\documentclass[universe,review,accept,oneauthor,pdftex,10pt,a4paper]{mdpi}
\pdfoutput=1

\firstpage{1}
\makeatletter
 \usepackage{xcolor}
\makeatother
\pubvolume{1}
\issuenum{1}
\articlenumber{0}
\pubyear{2024}
\copyrightyear{2024}

\history{Received: xxx; Accepted: xxx; Published: xxx}
\updates{yes}

\usepackage{amsfonts}
\usepackage{amssymb}
\usepackage{mathabx}
\usepackage{mathrsfs}
\usepackage[english]{babel}
\usepackage{url}
\usepackage[utf8]{inputenc}
\usepackage{geometry}
\usepackage{bm}

\newcommand{\ed}{\dot{\mathcal{E}}}

\newcommand{\beq}{\begin{equation}}
\newcommand{\eeq}{\end{equation}}

\Title{The modelling of  pulsar magnetosphere and radiation  }

\Author{Gang Cao$^{1}$, Xiongbang Yang$^{1}$, Li Zhang$^{2,\dagger}$}

\AuthorNames{Gan Cao,  Xiongbang, Yang, Li Zhang}

\address{%
$^{1}$ Department of Mathematics, Yunnan University of Finance and Economics, Kunming 650221, Yunnan,  China;  \\
$^{2}$ Department of Astronomy, Yunnan University, Key Laboratory of Astroparticle Physics of Yunnan Province, Kunming 650091, China; lizhang@ynu.edu.cn
        }

\abstract{
We review the recent advances in  the pulsar high-energy $\gamma$-ray observation and  the electrodynamics of the pulsar magnetospheres from the early vacuum model to the recent plasma-filled models by the numerical simulations. The numerical simulations have made the significant progresses toward the self-consistent modeling of the plasma-filled magnetosphere by including the particle acceleration and radiation. The current numerical simulations confirm a near force-free magnetosphere with the particle acceleration in the separatrix near the light cylinder and the current sheet outside the light cylinder, which can provide a good match to the recent high-energy $\gamma$-ray observations. The modeling of the combined multi-wavelength light curves, spectra, and polarization are expected to provide a stronger constrain on the geometry of the magnetic field lines, the location of the  particle acceleration and the emission region, and the emission mechanism in the pulsar magnetospheres.
}


\begin{document}

\section{Introduction}
\label{sec1}
Neutron stars are very fascinating astrophysical objects and excellent laboratories for studying the fundamental physics of the strong electromagnetic and gravitational fields. Pulsars are identified as rapidly rotating neutron stars with very strong surface magnetic fields ($10^8-10^{12}$ G) and rotation periods from milliseconds to seconds. It is believed that a pulsar will lose a substantial fraction of its rotational kinetic energy and convert it into particle acceleration and radiation. The periodic electromagnetic signals from these objects are visible across the entire electromagnetic spectrum from the radio to $\gamma$-ray bands. The multi-wavelength spectra from these objects are thought to be produced by the synchrotron, curvature, and inverse-Compton radiation from the high-energy particle accelerated by unscreened electric fields in the magnetosphere. However, it is still unclear where the particles  are accelerated in the magnetosphere and how the subsequent radiation is produced. Although the pulsars were first discovered in the radio band \citep{hew68}, it is difficult to understand the origin of the radio emission due to the coherent nature in the radio band. Since the emission at the higher energies originates from the incoherent process, the origin of the pulsar emission and the particle acceleration in the magnetosphere can  better be understood at the higher energies (for example, in the $\gamma$-ray band).
Large area telescope (LAT) of Fermi Gamma-Ray Space Telescope and ground-based Cherenkov telescopes have opened a new era in the study of the pulsar $\gamma$-ray physics. Fermi-LAT observations have discovered a large number of the $\gamma$-ray pulsars with the measurement of good-quality light curves and spectra in the GeV $\gamma$-ray bands \citep{abd10a,abd13,Smith23}.
Ground-based Cherenkov telescopes have  also detected the sub-TeV $\gamma$-ray emission  for a growing number of the Fermi pulsars \citep{ans16,abd18,spi19,acc20}.
The location and geometry of the particle acceleration and emission regions  will strongly imprint on the multi-wavelength light curves and the phase-resolved spectra. The observed multi-wavelength light curves and spectra can provide the powerful probes of the emission region geometry and emission mechanics. The properties of the pulsar multi-wavelength emission are directly associated to the structure of the  pulsar magnetospheres, which require to have a better and deeper understanding of the realistic pulsar magnetosphere.


In the pulsar magnetosphere, the electromagnetic field, the particle dynamics and the subsequent radiation mechanics are all coupled mutually. This requires us to perform a self-consistent calculation of the Maxwell equations, the particle dynamics, and the subsequent radiation to model the realistic pulsar magnetospheres. Significant progresses have been made towards the self-consistent modeling of the pulsar magnetosphere over the last decades. The first solution of the pulsar magnetosphere is the vacuum fields, which has the advantage of the analytical expression derived by Deutsch (1955) \citep{deu55}. However, the vacuum solution is not a real pulsar model, since it does not take into account the modification of the current and charge on the magnetosphere structure.  It is well established that the pulsar magnetosphere would be filled with plasmas created by the pair cascades.
A zeroth order approximation of the plasma-filled magnetosphere is referred to as the force-free electrodynamics, where enough particles are produced to  screen all
accelerating electric field so that force-free condition $\bm{E} \cdot \bm{B}=0$ holds everywhere in the magnetosphere.
The force solutions have recently became available by the numerical simulation  for the aligned rotator \citep{con99,kom06,mck06,yu11,par12,cao16a} and for the oblique rotator \citep{spi06,kal09,pet12a,eti17}. Although the force-free solutions are thought to be closest to realistic pulsar magnetospheres, they do not allow any particle acceleration and the production of radiation in the magnetospheres.  The resistive pulsar magnetosphere is then developed by introducing the self-consistent accelerating electric field  with a prescribed conductivity \citep{li12,kal12a,cao16b}. The resistive magnetospheres are still not self-consistent, since they can not provide microscopic physical information that produces the magnetospheric conductivity. The kinetic particle-in-cell (PIC) models are later developed to self-consistently compute the feedback between the particle motions, the radiating photons and the electromagnetic fields \citep{phi14,che14,bel15,cer15,phi15a,kal18,bra18}. The PIC models are not yet fully self-consistent, since they  can not use the real pulsar parameters to model the pulsar magnetosphere and predict the pulsar emission  due to large scale separation between the plasma frequency and the pulsar rotation frequency. A clever method with the combined force-free and Aristotelian electrodynamics (AE) are recently developed to model the pulsar magnetosphere \citep{con16a,pet20a,cao20,pet22}, which can include the back-reaction of the emitting photons onto particle dynamics and introduce the local accelerating electric field to produce the observed pulsar emission with the real pulsar parameters. All these magnetospheric simulations obtain a near force-free magnetosphere with the particle acceleration near the current sheet outside the light cylinder (LC). The good matches between the magnetospheric simulations and the $\gamma$-ray observations indicate that the particle acceleration and the $\gamma$-ray emission mainly originate from the regions near the current sheet outside the LC. However, there are some disagreements over whether the synchrotron or curvature radiation from the current sheet is the main Fermi $\gamma$-ray emission mechanism. The light curve and spectral information is not sufficient to distinguish between different emission region geometry and emission mechanisms in the pulsar magnetosphere.
The pulsar polarization information can provide another independent constraint on the emission region geometry and emission mechanisms. The modeling of the combined multi-wavelength emission and polarization are expected to provide a stronger constrain on the geometry of the magnetic field lines, the location of the  particle acceleration and the emission region, and the emission mechanism in the pulsar magnetospheres.


In this review, we summarize the recent advances  in  the pulsar high-energy $\gamma$-ray observations and the electrodynamics of the pulsar magnetospheres from the early vacuum model to  recent plasma-filled models by the numerical simulations. We focus on  the  numerical simulations of  recent plasma-filled  magnetospheres and  the fruitful feedbacks between the $\gamma$-ray observations and the magnetospheric simulations from the force-free, resistive, PIC and  combined force-free and AE models.
A similar excellent review was recently presented by Philippov \& Kramer (2022) \citep{phi22}, where they focus on the PIC simulations of the pulsar magnetospheres and the meaningful connections between the  radio and $\gamma$-ray observations and the PIC simulations. More recent reviews can be found in \citep{con16b,har16,pet16a,cer17,har22}.

\section{Pulsar gamma-ray observations}
\label{sec2}

\begin{figure*}
\center
\begin{tabular}{cccc}
\includegraphics[width=7.5 cm,height=9. cm]{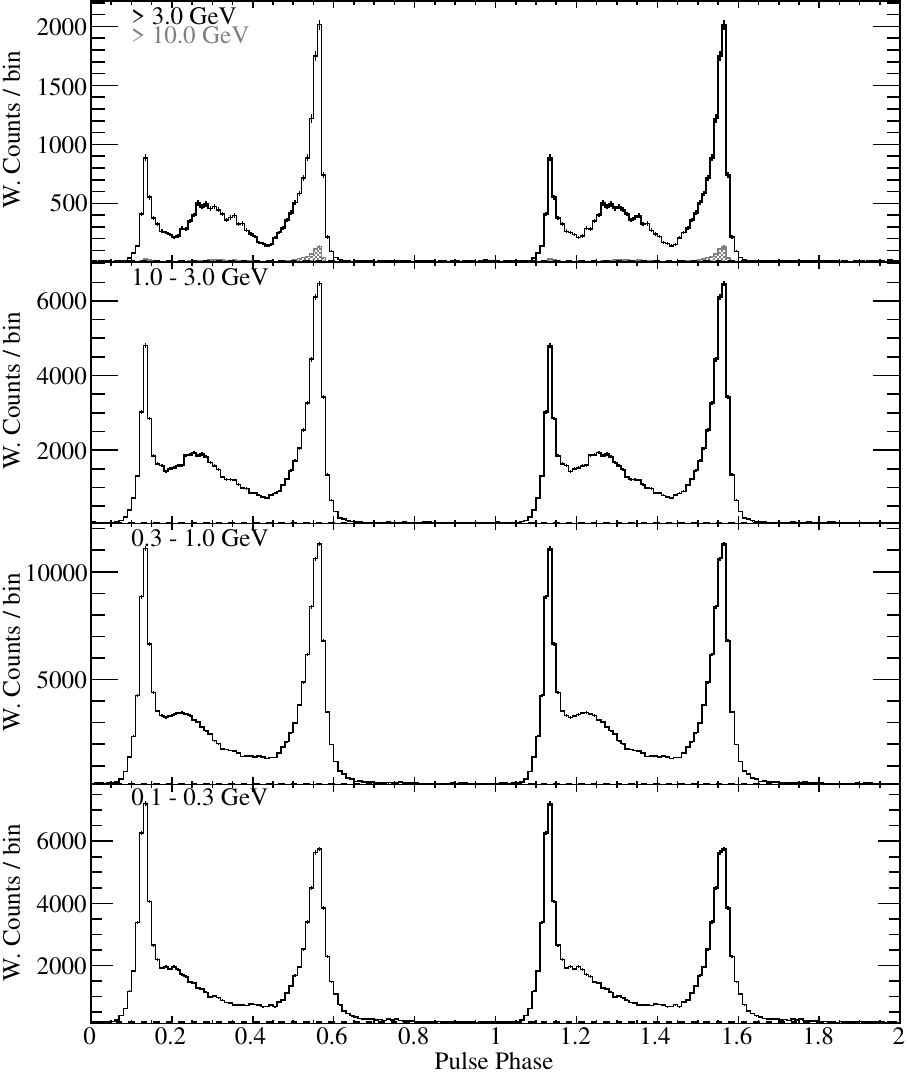}
\end{tabular}
\caption{The Fermi observed energy-dependent $\gamma$-ray light curves for the Vela pulsar. It is clearly seen  that the relative ratio of the first to the second $\gamma$-ray peak decreases with increasing energies. The figure is taken from Abdo et al. (2010b) \citep{abd10b}. }
\label{fig1}
\end{figure*}

Fermi Gamma-Ray Space Telescope launched in 2008 provides an unprecedented opportunity to study the physics of the pulsar $\gamma$-ray emission. Three Fermi Pulsar Catalogs have been published in \citep{abd10a,abd13,Smith23}. In fact, more than 40 $\gamma$-ray pulsars were detected in the first year of the Fermi large area telescope (LAT) operations \citep{abd10a}. 117 $\gamma$-ray pulsars are listed in the Second Fermi Pulsar Catalog \citep{abd13}. To date, Fermi-LAT have discovered 294 confirmed $\gamma$-ray pulsars, including 150 young gamma-ray pulsars and 144 millisecond pulsars (MSPs) \citep{Smith23}. The detected $\gamma$-ray pulsars can be divided into young radio-loud, young radio-quiet, and MSPs. Fermi observations have demonstrated that the pulsars are the dominant source of the GeV $\gamma$-rays in the Milky Way. A wealth of high-quality light curves, phase-averaged and phase-resolved spectra are now available. In the third Fermi Pulsar Catalog, the light curves show a wide variety of the pulsed profiles: single-peak, double-peak, and multi-peak profiles, where their ratio are $\sim 0.26$, $\sim 0.53$, and $\sim 0.21$, separately. Moreover, in 120 Fermi pulsars with the double-peak profiles, the distribution of the phase separations ($\Delta$) is not uniform, varying from $\sim 0.082$ to $\sim 0.73$, where  55\% Fermi pulsars have $\Delta\ge0.4$. On the other hand, the radio peaks of young Fermi pulsars typically lead the first $\gamma$-ray peak by a small fraction of rotation period, but it is not the case for Fermi MSPs. The double-peaked $\gamma$-ray light curves usually show an energy evolution where the relative ratio of the first to the second $\gamma$-ray peak decreases toward the higher energies (see Figure \ref{fig1}). The observed $\gamma$-ray spectra  are usually characterised by a power law  with an exponential cutoff, and the cutoff energy typically lies in the range of 1–5 GeV. The large sample of the Fermi $\gamma$-ray pulsars also find some  statistical trends and correlations among their  observational quantities, typically including the radio lag  and peak separation  correlation, as well as the dependence of the $\gamma$-ray luminosity on spin-down power.

\begin{figure*}
\center
\begin{tabular}{cccccccccccccc}
\includegraphics[width=6.5 cm,height=5.9 cm]{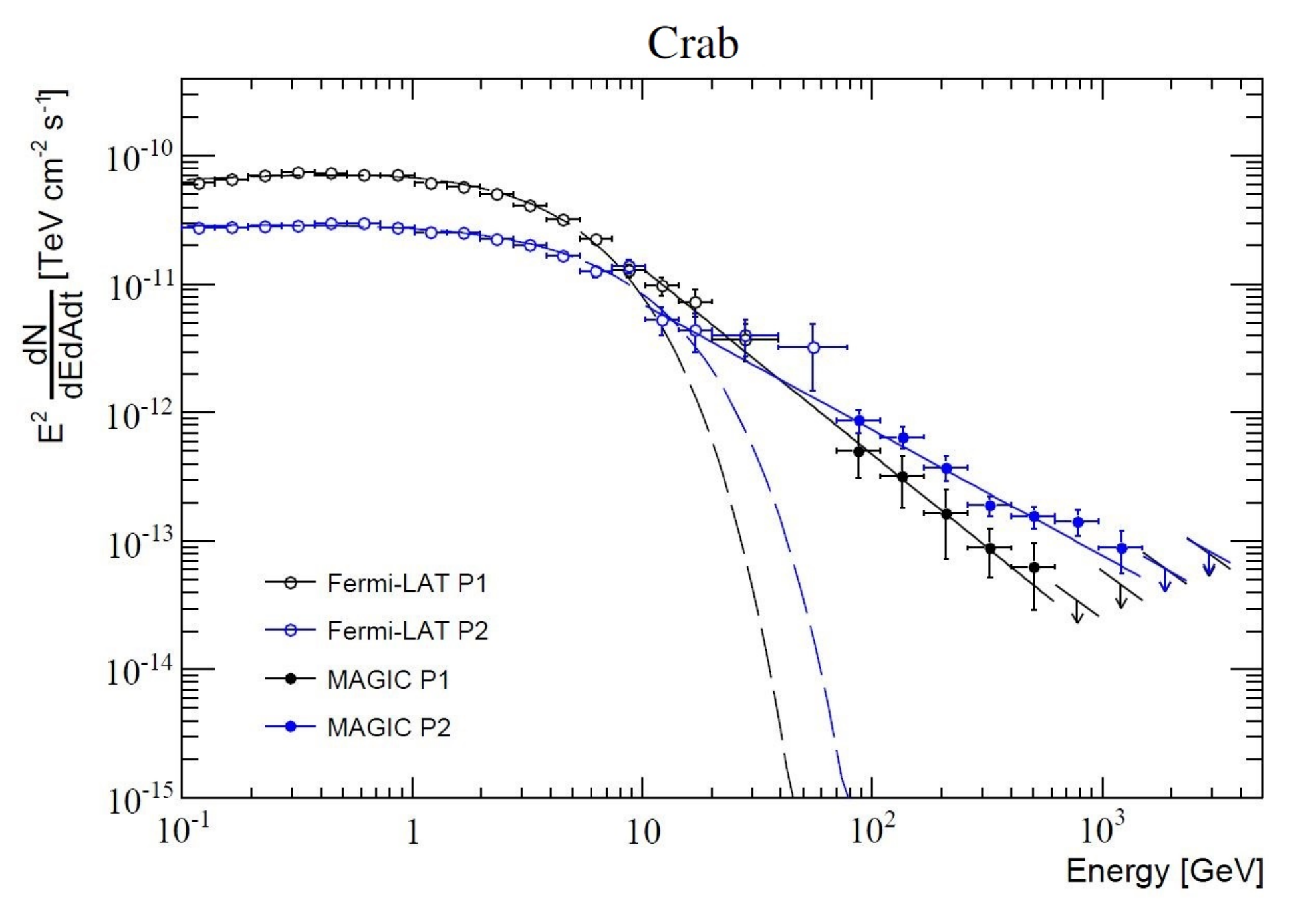} \qquad
\includegraphics[width=5.9 cm,height=5.75 cm]{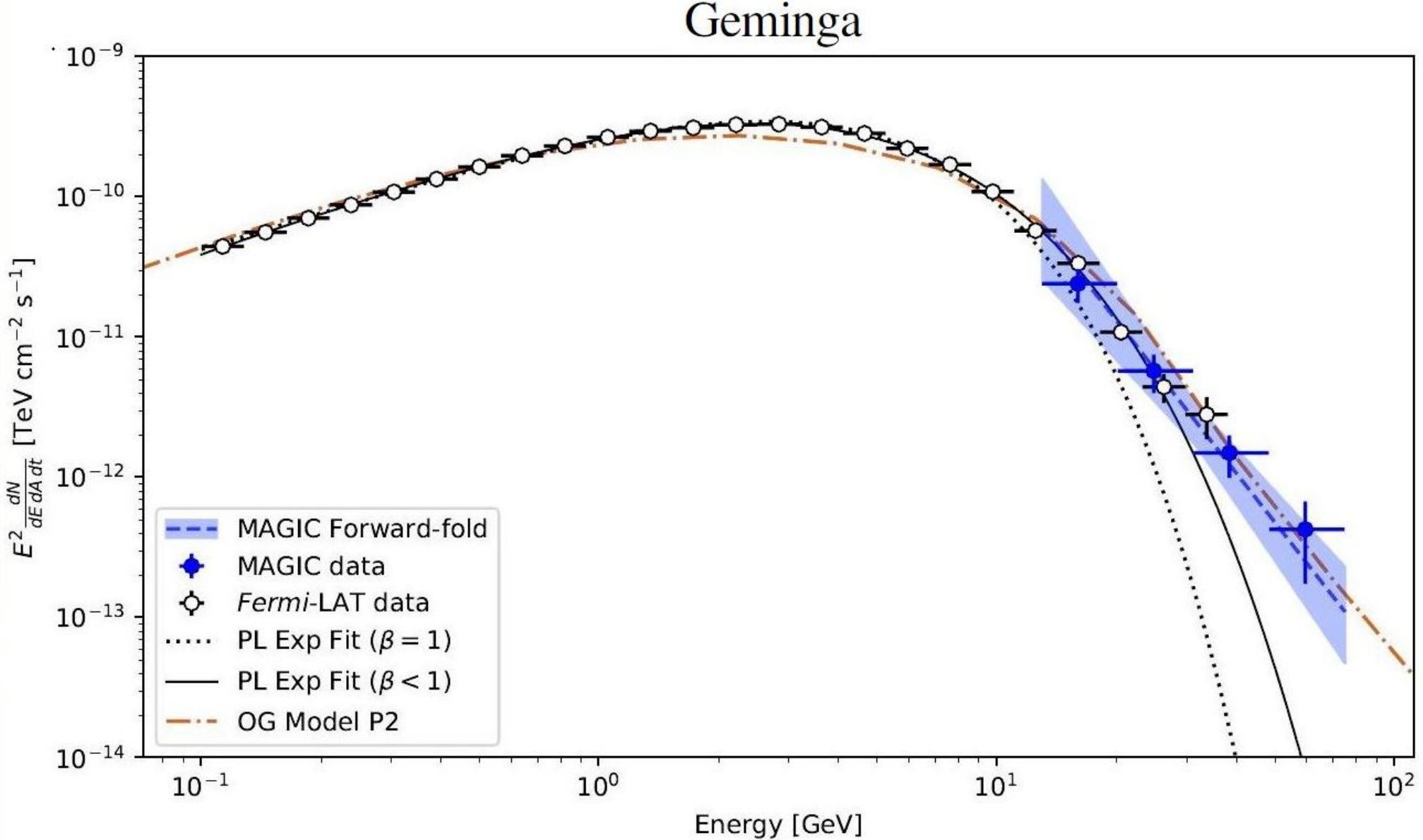}
\end{tabular}
\caption{The $\gamma$-ray spectra of the Geminga and Vela pulsar measured by the MAGIC telescopes and the Fermi-LAT. The figure is taken from Ansoldi et al. (2016) \citep{ans16} and Acciari et al. (2020) \citep{acc20}. }
\label{fig2}
\end{figure*}

The detections of the pulsed TeV  emission from the ground-based Cherenkov telescopes can provide some new clues on the  particle acceleration and emission mechanics in the pulsar magnetosphere. The ground-based Cherenkov telescopes have detected the pulsed $\gamma$-ray emission at the sub-TeV energies for  a growing number of Fermi pulsars. These pulsars include the Crab, Vela, Geminga, and PSR B1706-44 with the first two having pulsed emission above 1 TeV. MAGIC have detected the pulsed emission to up to 1.5~TeV from the Crab pulsar \citep{ans16}. H.E.S.S-II have reported the detection of the pulsed emission from the 20~GeV to 100~GeV and also above 3 TeV from Vela pulsar \citep{abd18}, and in the energy range below 100 GeV from PSR B1706-44 \citep{spi19}. MAGIC also have announced the detection of the pulsed emission between 15 GeV and 75 GeV  with the significance high up to 6.3 $\sigma$ from the Geminga pulsar \citep{acc20}. The observed light curves show only the second $\gamma$-ray peak, which confirms the energy evolution of the Fermi light curves with the decreasing ratio of the first to second  $\gamma$-ray  peaks toward the higher energies. The observed $\gamma$-ray spectrum is smoothly connected to the Fermi spectrum, which shows a  power-law extension of the Fermi spectrum for the Vela, Geminga, and PSR B1706-44 pulsars and a broken  power-law extension of the Fermi spectrum for the Crab pulsar (see Figure \ref{fig2}). The detected TeV photon energy directly measure the maximum energy of the accelerated particles, which can provide an excellent diagnostic of  the  particle acceleration and emission mechanics in the pulsar magnetospheres.

\section{Time-dependent Maxwell equations}
\label{sec3}

The pulsar magnetosphere can be obtained by numerically solving the time-dependent Maxwell equations
\begin{equation}
\begin{split}
{\partial {\bm B}\over \partial t}=-{\bm \nabla} \times {\bm E}\;,\\
{\partial  {\bm E}\over \partial t}={\bm \nabla} \times {\bm B}-{\bm J}\;,\\
\nabla\cdot{\bm B}=0\;,\\
\nabla\cdot{\bm E}=\rho_{\rm e}\;,
\end{split}
\label{Eq1}
\end{equation}
where ${\bm J}$ is the current density and $\rho_{\rm e}$ is the charge density.  The time-dependent Maxwell equations can be numerically solved by specifying the current density ${\bf{J}}$, which can be defined as a function of the fields (${\bm E}$, ${\bm B}$) in the MHD models or be obtained by solving the Vlasov equation in the PIC models.

The magnetospheric simulation is set up with an initialized dipole magnetic field in vacuum, whose magnetic moment $\mu$ is inclined at an angle $\alpha$ with respect to the rotation axis. The
neutron star is well approximated  by a perfect conductor, the inner boundary condition at the stellar surface is enforced  with a rotating electric field
\begin{eqnarray}
{\bm{E}} = -( {\bm \Omega } \times {\bf r} ) \times {\bm B}.
\label{Eq2}
\end{eqnarray}
A non-reflecting or absorbing outer boundary condition is imposed to prevent  artificial reflection from the outer boundary. The pulsar magnetosphere can be obtained by relaxing the time-dependent Maxwell equations towards a stationary state.

\section{The vacuum dipole magnetospheres}
\label{sec4}
\subsection{Field structure of  the vacuum magnetospheres}
The simplest model of the pulsar magnetosphere is vacuum dipole magnetosphere. It is assumed that the neutron star is surrounded by vacuum. There is no any plasma or particles outside the neutron star, which implies   $\rho_{\rm e}=0$ and  ${\bf J}=0$ in the time-dependent Maxwell equations. Therefore, the time-dependent Maxwell equations (\ref{Eq1}) are linear in the vacuum case, which can be analytically solved by using the inner boundary condition (\ref{Eq2}) at the stellar surface. 
Deutsch (1955)\citep{deu55} first derived the exact analytical expression for a vacuum dipole magnetosphere with a finite stellar radius. The general-relativistic extension of the Deutsch vacuum solution  including the space curvature and frame-dragging effects is also derived by \citep{pet17}. The exact analytical expression for the Deutsch vacuum solution is given by \cite{deu55,mic99,pet12a}.
\begin{equation}
\begin{split}
B_{r}(\bm{r},t) &= 2B \left[ \frac{R^3}{r^3}\cos\alpha \cos\theta+\frac{R}{r} \frac{h^{(1)}_{1}(k\,r)}{h^{(1)}_{1}(k\,R)}\sin\alpha  \sin\theta \, {\rm e}^{i \psi}  \right],  \\
B_{\theta}(\bm{r},t) &= B \left[ \frac{R^3}{r^3}\cos\alpha  \sin\theta+ \left(\frac{R}{r} \frac{\frac{d}{dr}\left(r\, h^{(1)}_{1}(k\,r)\right)}{h^{(1)}_{1}(k\,R)}
+\frac{R^2}{r^2_{\rm L}} \frac{h^{(1)}_{2}(k\,r)}{\frac{d}{dr}\left(r \, h^{(1)}_{2}(k\,r)\right)\mid_{R}} \right)\sin\alpha  \cos\theta \, {\rm e}^{i \psi} \right],               \\
B_{\phi}(\bm{r},t) &= B \left[  \frac{R}{r} \frac{\frac{d}{dr}\,\left(r \, h^{(1)}_{1}(k\,r)\right)}{h^{(1)}_{1}(k\,R)}
+\frac{R^2}{r^2_{\rm L}} \frac{h^{(1)}_{2}(k\,r)}{\frac{d}{dr}\,\left(r \, h^{(1)}_{2}(k\,r)\right)\mid_{R}}  \cos2\theta  \right]i\,\sin\alpha \, {\rm e}^{i\psi} .        \\
E_{r}(\bm{r},t) &= \frac{\Omega \, B \, R}{c} \left[ -\frac{R^4}{r^4}(3\cos^2\theta-1)\cos\alpha
+3 \sin\alpha  \sin2\theta \, {{\rm e}^{i\psi}} \frac{R}{r} \frac{h^{(1)}_{2}(k\,r)}{\frac{d}{dr}\left(r \, h^{(1)}_{2}(k\,r)\right)\mid_{R}}   \right] ,\\
E_{\theta}(\bm{r},t) &= \frac{\Omega \, B \, R}{c} \left[ \sin\alpha\,{{\rm e}^{i\psi}} \left(\frac{R}{r} \frac{\frac{d}{dr}\left(r \, h^{(1)}_{2}(k\,r)\right)}{\frac{d}{dr}\left(r \, h^{(1)}_{2}(k\,r)\right)\mid_{R}}\cos2\theta -\frac{h^{(1)}_{1}(k\,r)}{h^{(1)}_{1}(k\,R)} \right)  -\frac{R^4}{r^4}\sin2\theta\cos\alpha  \right], \\
E_{\phi}(\bm{r},t) &=\frac{\Omega \, B \, R}{c} \left[  \frac{R}{r} \frac{\frac{d}{dr}\left(r \, h^{(1)}_{2}(k\,r)\right)}{\frac{d}{dr}\left(r \, h^{(1)}_{2}(k\,r)\right)\mid_{R}} -\frac{h^{(1)}_{1}(k\,r)}{h^{(1)}_{1}(k\,R)} \right]i\,\sin\alpha \cos\theta \, {\rm e}^{i\psi},         \\
\end{split}
\label{Eq3}
\end{equation}
where $B$, $\Omega$ and $R$ are the surface magnetic field, the rotation speed, and the stellar radius, respectively, $h^{(1)}_{l}$ is the spherical Hankel functions of order $l$, $\psi=\phi-\Omega \, t$ is the instantaneous phase at time $t$, $k=1/r_{\rm L}$ is the wavenumber, and $r_{\rm L}$=c/$\Omega$ is the light cylinder. The physical solution is obtained by taking the real parts of each component.

\begin{figure*}
\center
\begin{tabular}{cccccccccccccc}
\includegraphics[width=7 cm,height=7 cm]{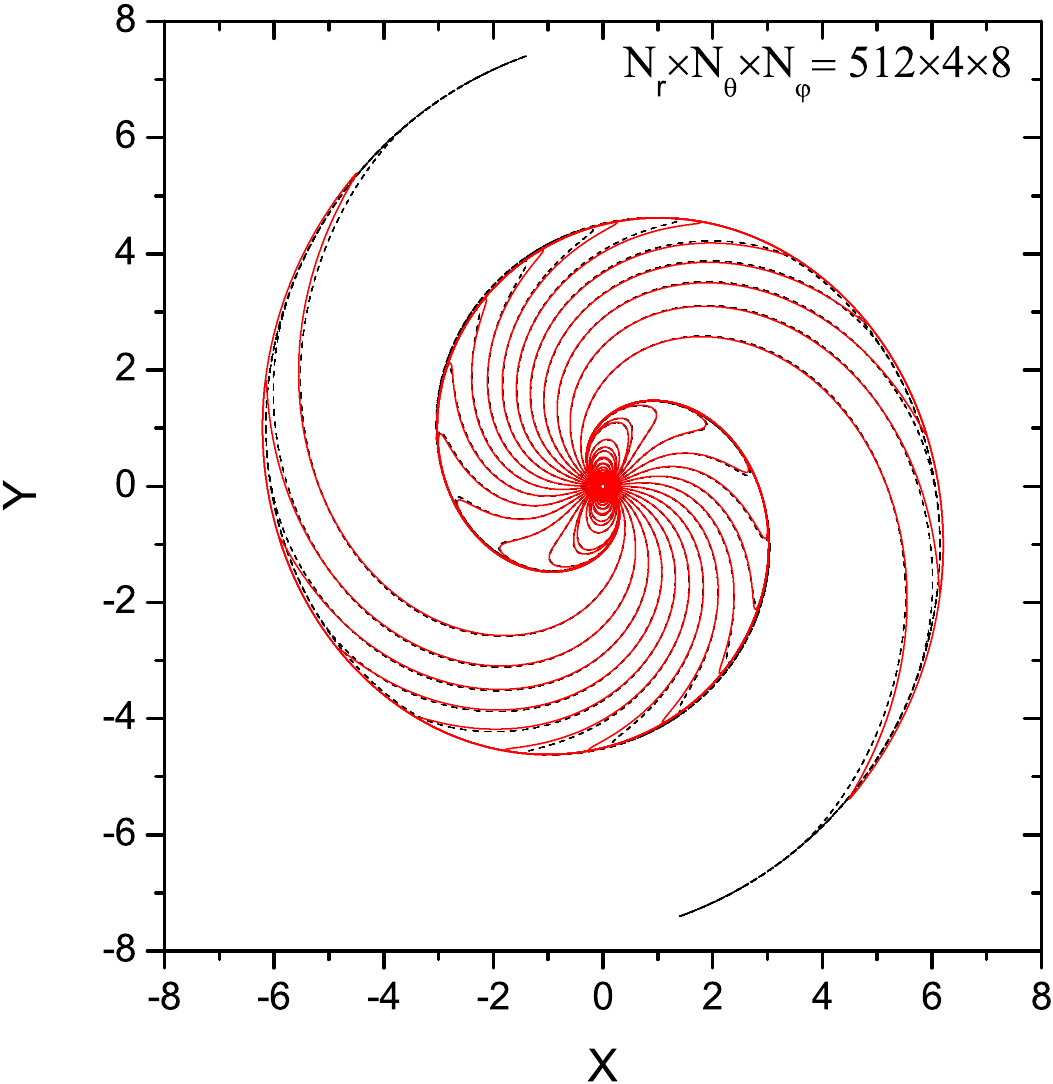}
\end{tabular}
\caption{Magnetic field lines of the perpendicular Deutsch vacuum field in the equatorial plane. The numerical and analytical solutions are shown as the black dashed and red solid lines, respectively. The figure is taken from Cao et al. (2016a) \citep{cao16a}.}
\label{fig3}
\end{figure*}

The Deutsch vacuum solution can be used as a benchmark to investigate plasma-filled pulsar magnetosphere by including the plasma feedback onto the vacuum fields. It can be also served as a reference solution for checking the algorithm and accuracy of the numerical solution. A spectral method was  developed by P\'{e}tri (2012a) \citep{pet12a} and Cao et al. (2016a) \citep{cao16a} to compute the structure of pulsar magnetosphere, they tested the spectral algorithm by comparing the numerical solution and the Deutsch vacuum solution. Figure \ref{fig3} shows the magnetic field lines of the perpendicular vacuum rotator in the equatorial plane for the numerical solution and analytical solutions. It is shown that the spectral method can accurately compute  vacuum electromagnetic fields.
The Deutsch vacuum solution is very close to the static dipole near the stellar surface, but the sweepback of the field lines produce a significant toroidal magnetic field by the displacement currents outside the LC. The rotating magnetic field induces a very strong quadrupole electric field near the stellar surface, which falls off as $1/r^4$ much faster than the magnetic field $1/r^3$. The sweepback of magnetic field lines distorts the polar cap  and shifts it towards the trailing side. It is well believed that the pulsars lose their rotational kinetic energy by particle acceleration and radiation, leading to the spin down of the pulsar. A rotating vacuum dipole  radiates an outward electromagnetic wave by magnetic dipole radiation, which is proposed to explain the  luminosity of the Crab nebula \citep{pac67}. The Poynting flux for a rotating vacuum dipole is given by
\begin{eqnarray}
L_{\rm vac}=\frac{2}{3} \frac{\mu^2\,\Omega^4}{c^3}\, {\rm sin}^2\alpha .
\label{Eq4}
\end{eqnarray}
It is found that the vacuum aligned  rotator is not an active pulsar magnetosphere, $L_{\rm alinged,vac}=0$, which does not dissipate the Poynting flux and hence does not spin down.

\subsection{The radiation models from the vacuum magnetospheres}

The vacuum dipole field is the first solution of the global pulsar magnetosphere, which has the advantage of  the analytical expression given by equation (\ref{Eq3}). Therefore, it is widely used as the  background magnetic field of the pulsar radiation modeling in the early study of the pulsar emission, even though Goldreich \& Julian (1969) \citep{gol69} pointed out that the realistic pulsar magnetosphere can not be surrounded by vacuum. Some acceleration gap models are developed to explain the observed pulsar emission based on the vacuum dipole field. These gap models usually use the  vacuum dipole magnetosphere as an approximation to the force-free magnetosphere. A local deviation with the force-free condition forms an acceleration gap with an unscreened accelerating electric field, which accelerates the particles and produce the observed pulsed emission. The accelerating electric field, $\bm{E}_{\|}$, is found by solving the Poisson's equation
\begin{eqnarray}
\nabla\cdot\bm{E}_{\|}=4\pi(\rho-\rho_{\rm GJ}),
\label{Eq5}
\end{eqnarray}
where $\rho$ is the charge density and $\rho_{\rm GJ}$ is the Goldreich–Julian charge density, which is minimum density required to locally screen the accelerating electric field.
However, the accelerating electric fields in these gap models do not self-consistently come from a global magnetosphere model.

\begin{figure*}
\center
\begin{tabular}{cccccccccccccc}
\includegraphics[width=9. cm,height=8. cm]{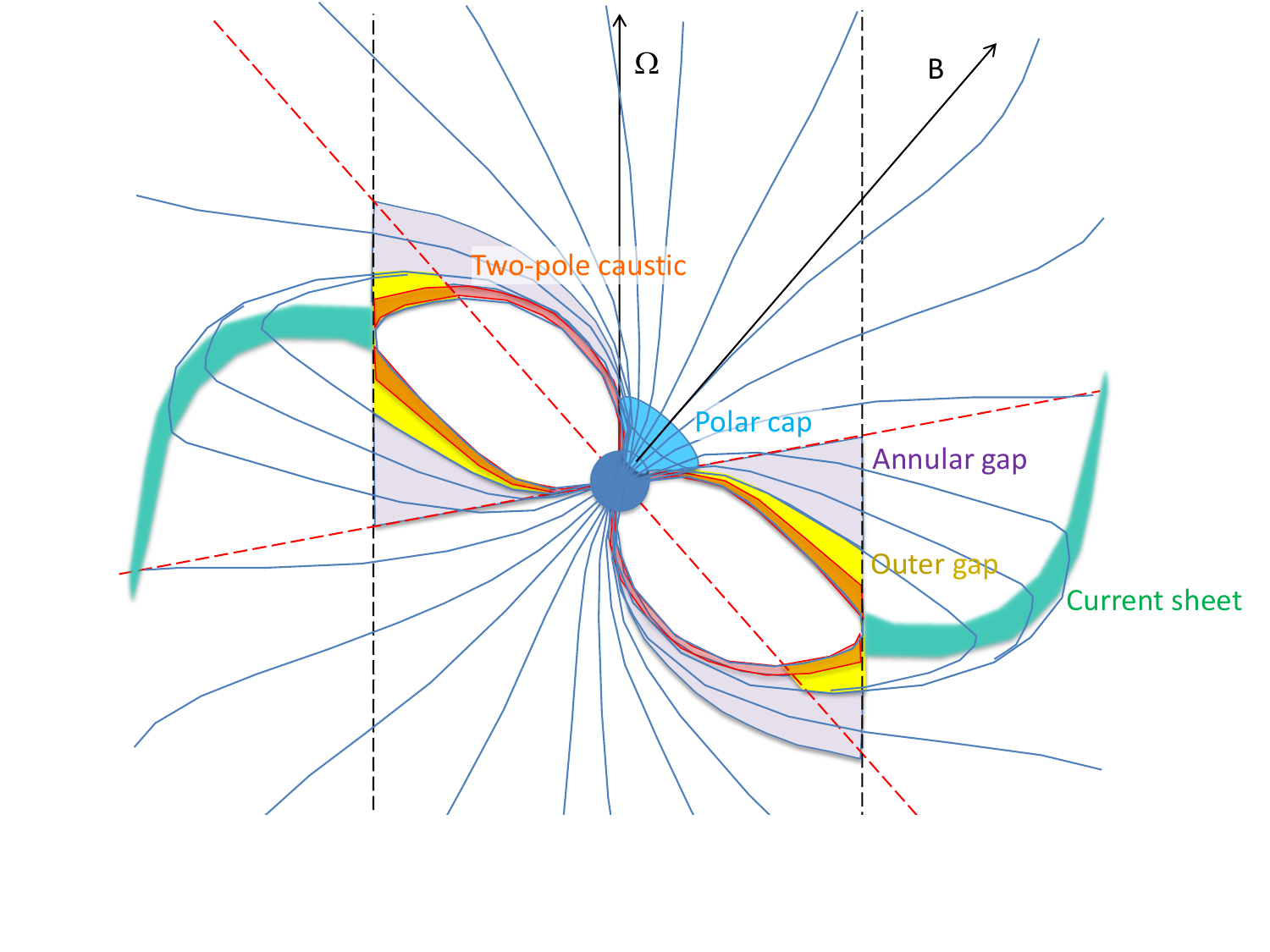}
\end{tabular}
\caption{The schematic diagram of different radiation models. The figure is taken from Harding, A. K (2022) \citep{har22}.}
\label{fig4}
\end{figure*}

The acceleration and radiation of the particles in the gap models are generally assumed to occur inside the LC. Such gap models have been developed to model the pulsar emission phenomena, including the polar gap (PC) \citep{rud75,dau82}, slot gap (SG, or the two-pole caustic model, TPC) \citep{dyk03,mus04} and outer gap (OG) \citep{che86,zha97,che00} models. The PC model locates the particle acceleration at  about several stellar radius above the PC surface. It is assumed that the particles are freely extracted from the neutron star surface, which leaves a charge deficit to produce an $\bm{E}_{\|}$ above the PC surface. These particles are accelerated by the  $\bm{E}_{\|}$ to radiate the $\gamma$-ray photons by the curvature radiation. The produced $\gamma$-ray photons induce a pair cascade by one-photon magnetic ($\gamma$$-$B) pair production in the strong magnetic field.  The pair cascades produce a low-altitude pair formation front, above which all  $\bm{E}_{\|}$ are screened by the pairs. If the pair formation front is extended to  the high latitude, a SG gap is formed between the last closed lines and the pair formation front. These particles are not accelerated to  enough high energy  to produce  the pairs at the low altitude, because the $\bm{E}_{\|}$ near the last open field line is weak. However, they can be continuously accelerated  to high altitude and produce the  $\gamma$-ray photons by the curvature or  inverse-Compton radiation. The OG model is a vacuum gap  between the null-charge surfaces and the last closed field lines, where an $\bm{E}_{\|}$  are developed because of charge depletion. This gap accelerates the particles to produce the $\gamma$-ray photons by the curvature radiation. These $\gamma$-ray photons can interact with the surface X-ray photons to initiate a pair cascade by the photon-photon ($\gamma-\gamma$) interaction. The size of the gap is limited by the screening of the $\bm{E}_{\|}$ by the pair cascades. The OG electrodynamical simulations also show that the inner boundary of the outer gap can shift toward the stellar surface from the null-charge surface when the current through the gap is taken into account \citep{hir03,tak04,tak06}. The current sheet or the stripped wind (SW) model is also proposed as a potential sites of the particle acceleration in addition to the traditional gap regions \citep{cor90,pet05,pet09,pet12b,uzd14,cz19}, where the particle acceleration and high-energy radiation is located in the current sheet near or outside the LC. Beside, an annular gap model is also developed to model the pulsar emission \citep{qia04,qia07}, which locates the emission region  between the critical and last open field lines. The schematic diagram of different radiation models are depicted in Figure \ref{fig4}.

\begin{figure*}
\center
\begin{tabular}{cccccccccccccc}
\includegraphics[width=11. cm,height=10.5 cm]{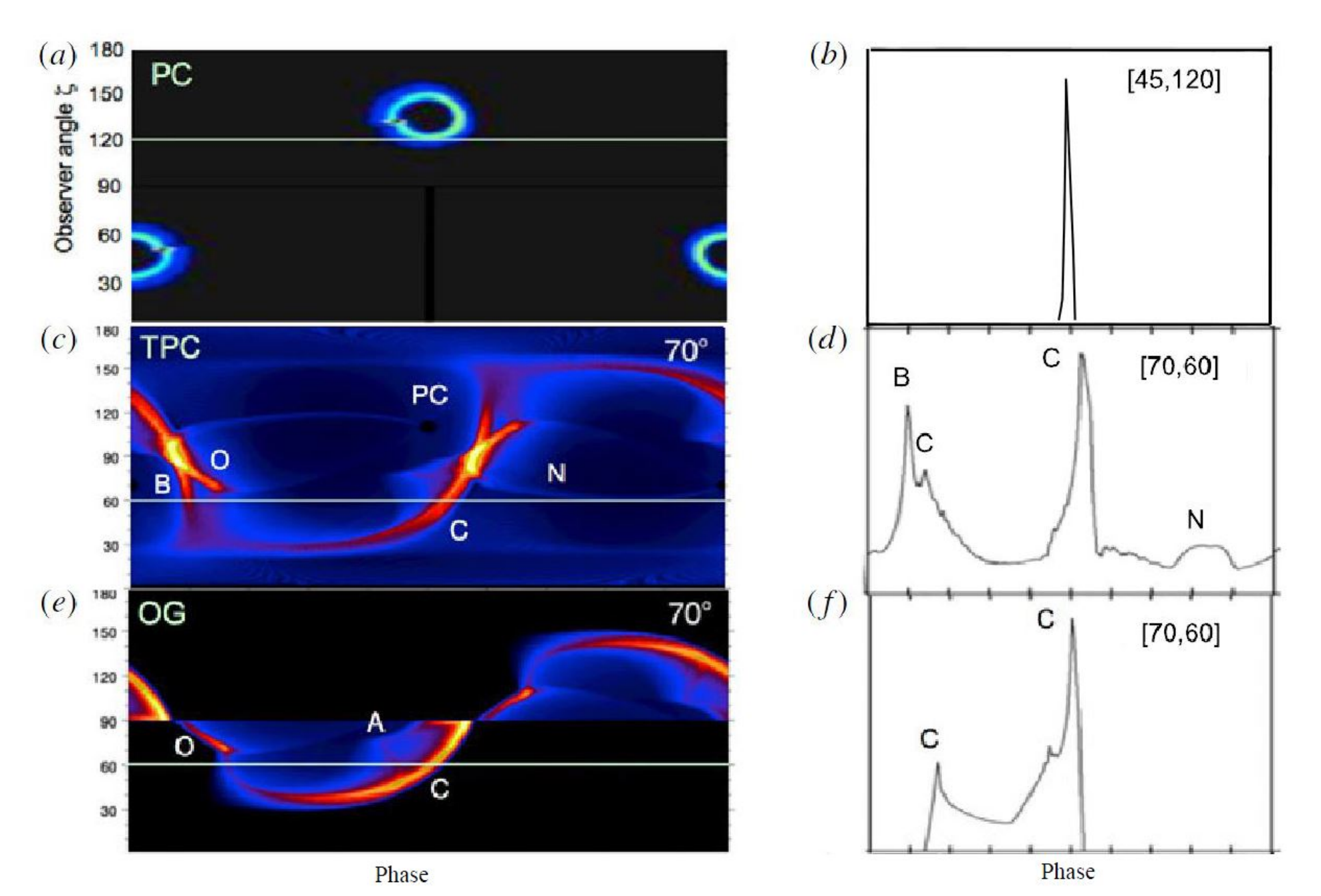}
\end{tabular}
\caption{Sky maps and light curves from the PC, SG and OG model at different inclination angles and viewing angles. The figure is taken from Harding, A. K (2016) \citep{har16}. }
\label{fig5}
\end{figure*}

\subsection{Light curve and spectra modelling from the vacuum magnetospheres}

The Fermi $\gamma$-ray observations have accumulated a wealth of the high-quality pulsar light curves over the last decades. A geometric light curve model is usually used to fit the pulsar Fermi light curves based on the PC, SG and OG models with the vacuum dipole field, which can help us to test different emission models and constrain the geometry of different emission regions.
The geometric light curve can be produced by assuming the uniform emissivity along the field lines in the comoving frame. The direction of the photon emission can be obtained by a revised aberration formula given by \citep{bai10a}. The bright caustic is a common feature of the 3D emission pattern, which is formed when the phase delays from the field line sweepback cancels those due to the aberration and time-delay effect. The emission from different field lines arrives to an observer in the same phase. One or two sharp peaks in the light curves are seen when the viewing angle of the observer crosses the caustic. Figure \ref{fig5} shows the sky maps and light curves from the PC, SG, and OG model at different inclination angles and  viewing angles. The PC model predicts a  hollow-cone emission pattern with the peak at phase 0 and 0.5, which produces a near phase alignment of the radio and gamma-ray profiles. These models have difficulty in producing  the widely separated $\gamma$-ray light curves seen by Fermi-LAT. The original OG model can produce the double-peak light curve profiles with little off-peak emission from only one magnetic pole, because the gap forms above the null-charge surface where $\zeta=90^{\circ}$ (but, the revised version of the OG model can solve the problem \citep[e.g.,]{LZJ13}). The SG model also can produce the double-peak light curve profiles but from the opposite poles. Compared with the original OG model, the SG model has the emission below and above null-charge surface, which can produce a significant off-peak component seen in some young radio-faint pulsars and MSPs.

The geometric light curve models based on different emission models were used to fit a sample of the  $\gamma$-ray pulsar light curves in the 2PC \citep{wat09,rom10}.  It is found that the OG and SG models can generally fit the observed $\gamma$-ray light curves. These studies can constrain the  magnetosphere geometry of each pulsar described by magnetic inclination angles $\alpha$  and viewing angles $\zeta$. The more constraints can be obtained by jointly fitting the pulsar radio and $\gamma$-ray light curves \citep{ven09,ven12,joh14,pie15,pie16}.
The geometric light curve model cannot give the  information of the pulsar energetics and spectra. The modeling of pulsar energy-dependent light curves and spectra can provide a better constraint on the modeling parameters and radiation mechanisms. The PC model predicts a super-exponential spectral shape due to magnetic pair production near the neutron star surface. The measurement of
the exponential cutoff spectra by Fermi observation directly ruled out the super-exponential spectral shape  predicted by the PC models. Therefore, the SG and OG models are more favored, since these models  predict a exponential cutoff curvature spectrum at several GeV energies in agreement with the Fermi observation.  The energy-dependent  $\gamma$-ray light curves and spectra of the Vela pulsar were modeled by the curvature radiation for the OG \citep{wan11} and  annular gap model \citep{du11}. The multi-wavelength radiation of the Crab pulsar was also modeled by the OG \citep{li10}, SG \citep{har08} and  annular gap \citep{du12} models. Moreover, the pulsar polarization properties were  explored by the OG \citep{dyk04} and SG models \citep{tak07}. These models generally predict a fast swing of the position angle (PA) and the depolarization dips during the pulse peaks, which are similar to the measured polarization of the Crab optical emission. It is noted that the pulsar emission cannot be modeled beyond the LC in these gap models, since the particle trajectory does not follow the magnetic line beyond the LC where the particle velocity exceeds the speed of light. 

\begin{figure*}
\center
\begin{tabular}{cccccccccccccc}
\includegraphics[width=16.5 cm,height=6. cm]{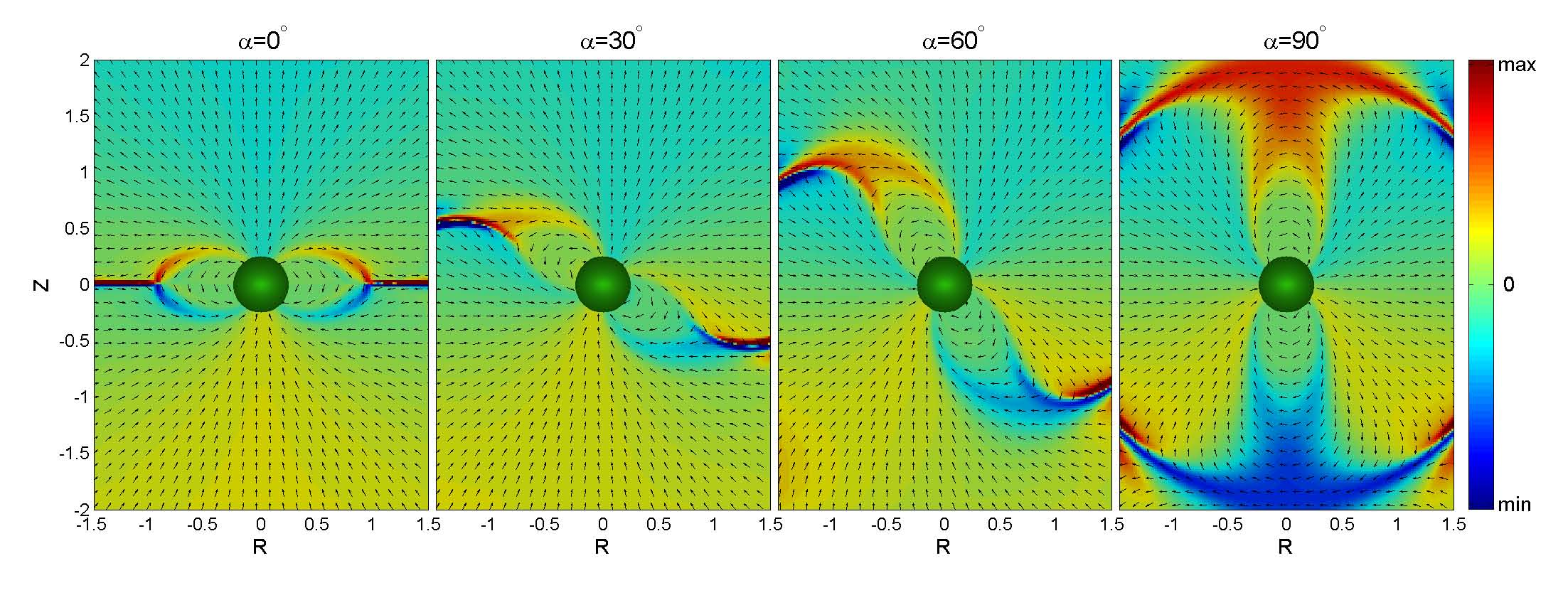}
\end{tabular}
\caption{The  distribution of the force-free magnetic field line and the current density parallel to the magnetic field  for a range of the inclination angles. The figure is taken from Bai \& Spitkovsky  (2010b) \citep{bai10b}.}
\label{fig6}
\end{figure*}

\section{The Force-free magnetospheres}
\label{sec5}

\subsection{Field structure of the force-free magnetospheres}

After the discovery of the Deutsch vacuum solution, Goldreich \& Julian (1969) \citep{gol69} realized that the Deutsch vacuum solution produces an $\bm{E}_{\|}$ accelerating electric field along the magnetic line at the stellar surface. This accelerating electric field lifts the particles off the star surface to fill the magnetosphere. On the other hand, these lifted particles are accelerated by the $\bm{E}_{\|}$ to radiate the $\gamma$-ray photons by the curvature radiation and inverse-Compton scattering of the surface thermal X-ray photons, which initiates a pair cascade by the $\gamma$$-$B interaction near the stellar surface and $\gamma-\gamma$ interaction near the LC. The pair cascades produce more secondary pairs to fill the magnetosphere. If the plasma density is much higher than the GJ density, all  accelerating electric fields are shorted out so that the force-free condition $\bm{E}\cdot\bm{B}=0$ holds in the magnetosphere everywhere. This corresponds to the zeroth-order approximation of the plasma-filled magnetosphere and referred to as force-free electrodynamic. Observations of pulsar wind nebulae also indicate that a large of plasma are presented in the magnetosphere \citep{buc11}. Therefore, realistic pulsar magnetosphere should  be closer to the force-free field than the vacuum dipole field.

The force-free approximations assume the negligible plasma inertia and pressure, the Lorentz force acting on the plasma fluid element will vanish,
\begin{eqnarray}
\bm{\rho_e}\bm{E}+\bm{J}\times\bm{B}=0.
\label{Eq6}
\end{eqnarray}
By taking the cross product to Equation (\ref{Eq6}) with $\bm{B}$, we have
\begin{eqnarray}
\bm{J}=\bm{\rho_e}\frac{\bm{E}\times\bm{B}}{B^2}+\frac{(\bm{J}\cdot\bm{B})\bm{B}}{B^2}=0.
\label{Eq7}
\end{eqnarray}
where $B^2\neq0$ have been assumed. By using the force-free condition $\bm{E}\cdot\bm{B}=0$, we have
\begin{eqnarray}
\frac{\partial}{\partial t}(\bm{E}\cdot\bm{B})=\frac{\partial \bm{E}}{\partial t}\cdot\bm{B}+\bm{E}\cdot\frac{\partial\bm{B}}{\partial t}=0.
\label{Eq8}
\end{eqnarray}
By substituting the time-derivative terms of $\bm{E}$ and $\bm{B}$ with the time-dependent Maxwell equation (\ref{Eq1}), Equation (\ref{Eq8}) becomes
\begin{eqnarray}
( {\bf \nabla} \times {\bf B} )\cdot\bm{B}-\bm{J}\cdot\bm{B}-\bm{E}\cdot( {\bf \nabla} \times {\bf B} )=0.
\label{Eq9}
\end{eqnarray}
By substituting $\bm{J}\cdot\bm{B}$ term in the Equation (\ref{Eq7}) into Equation (\ref{Eq9}), the force-free current density is given by \citep{gru99,bla02,kat17}
\begin{eqnarray}
\bm{J}=\bm{\rho_e}\frac{\bm{E}\times\bm{B}}{B^2}+\frac{( \bm{B}\cdot{\bf \nabla} \times {\bf B} +\bm{E}\cdot{\bf \nabla} \times {\bf B} )\bm{B}}{B^2}.
\label{Eq10}
\end{eqnarray}

\begin{figure*}
\center
\begin{tabular}{cccccccccccccc}
\includegraphics[width=16. cm,height=5.8 cm]{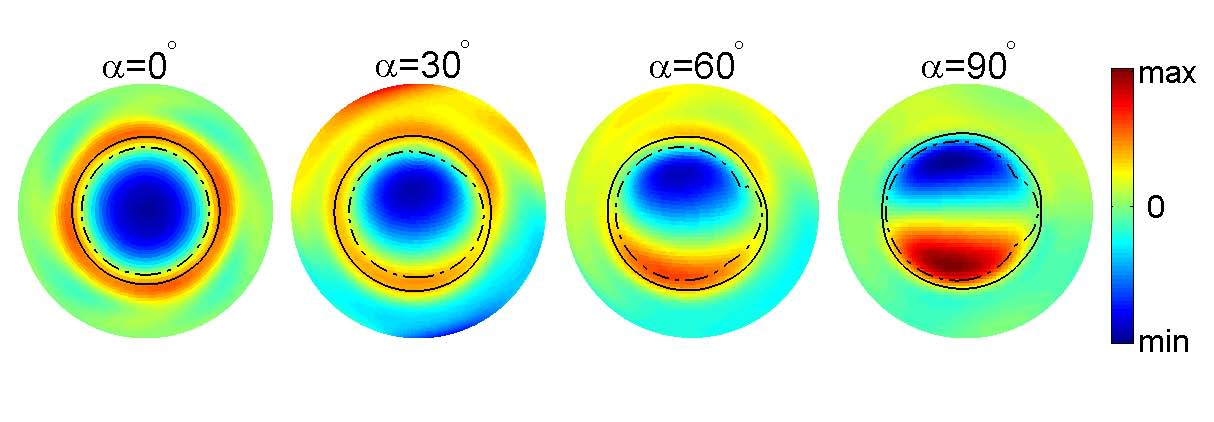}
\end{tabular}
\caption{Distribution of the current density parallel to the magnetic field on the force-free PCs (the solid lines) for a range of the inclination angles. The PC shapes for the vacuum fileld are also shown as the dash-dotted lines. The figure is taken from Bai \& Spitkovsky  (2010b) \citep{bai10b}. }
\label{fig7}
\end{figure*}


The time-dependent force-free equation (\ref{Eq1}) and (\ref{Eq10}) can be reduced to a one-dimensional pulsar equation  for  an axisymmetric aligned rotator \citep{sch73}. An exact analytical solution to the pulsar equation was found for a rotating magnetic monopole \citep{mic73}, which gives the far-field solution with the asymptotically radial field lines. However, no analytical solution is found for a rotating dipole. It was realized  that the force-free pulsar magnetospheres cannot be analytically solved  due to the problem of nonlinearity and singularity. This  renews new interest in the numerical simulation of pulsar magnetosphere thanks to the development of numerical techniques and increasing computer power. The numerical solution to the pulsar equation was first obtained by Contopoulos, et. al (1999, hereafter CKF) \citep{con99}, who  used a clever iterative algorithm to determine the poloidal current distribution that allows the the magnetic field lines smoothly cross the LC. The CKF solution consists of  the closed and  field line region separated by a Y-shaped current sheet outside the the light cylinder. It was later found that steady-state solutions can be  constructed with the Y-point inside the light cylinder\citep{goo04,tim06}. Therefore, it is compulsory for performing the time-dependent force-free simulation to solve the problem of the uniqueness and stability of the CKF solution. The time-dependent force-free equations for the oblique rotator was first solved by Spitkovsky (2006) \citep{spi06} through the finite-difference time-domain (FDTD) method in the Cartesian coordinate and followed by Kalapotharakos \& Contopoulos (2009) \citep{kal09}  with the incorporation of the absorbing outer boundary in the FDTD method. A 3D spectral method in the spherical coordinate was also developed by P\'{e}tri (2012) \citep{pet12a} and Cao et al. (2016b) \citep{cao16b} to solve the time-dependent force-free equations, which can impose the exact boundary condition by the use of the spherical coordinate. Moreover, the force-free electrodynamic was also extended to the full magnetohydrodynamic regime that takes plasma  inertial and pressures into account \citep{kom06,tch13} and  the general-relativistic force-free regime that takes the  frame-dragging and time–space curvature effects into account\citep{pac13,pet16b,car18}. All these time-dependent force-free simulations confirmed the closed–open CKF solution separated by a separatrix near the LC and a current sheet outside the the LC.

Figure \ref{fig6} shows the distributions of the force-free magnetic field line and the current density parallel to the magnetic field for different magnetic inclination angles. The presence of charges and currents modifies the field structure of the pulsar magnetosphere. This effect produces a toroidal magnetic field which increase a fraction of open field lines across the LC. The magnetic field  is close to the vacuum dipole with a subdominant toroidal magnetic field component inside the LC.  The magnetic field becomes  gradually dominated by the toroidal component and become asymptotically monopolar outside the LC. The extra toroidal component from the current increases the sweepback of the field lines  compared to the vacuum dipole.
A generic feature of the force-free magnetosphere is the formation of the equatorial current sheet outside the LC, where the magnetic field changes polarity and decreases to zero.
For the aligned rotator, the current sheet is axisymmetric along the rotational equatorial plane  outside the LC. For the oblique rotator, the current sheet becomes asymmetric with the undulating shape of the striped wind, which oscillates about the rotational equatorial plane  with the angular amplitude $2\alpha$ and a wavelength of $2\pi r_{\rm L}$.

Figure \ref{fig7} shows the distributions of the current density parallel to the magnetic field on the force-free polar caps (the solid lines) for different magnetic inclination angles.
The current flows out of one region of the polar cap and return to  another region of the polar cap. For  small $\alpha$,
the current flows out along the open field lines and returns to the star surface by the Y-shaped current sheet. As $\alpha$ increases, a gradually smaller fraction of the return current reaches the stellar surface  by the Y-shaped current sheet. For $\alpha=90^{\circ}$, all the current reaching the stellar surface is distributed over half of the polar caps, implying that the equatorial current sheet outside the LC is not directly connected to the star surface.
It is also shown that the force-free field has a larger PC size than that of the vacuum fields  because of the larger open field volume. Moreover, the force-free PC rim is more shifted toward the trailing side due to the higher field line sweepback relative to the vacuum field.
The Poynting flux for a  force-free rotator can be well approximated by
\begin{eqnarray}
L_{\rm FF}=\frac{\mu^2\,\Omega^4}{c^3}\,(1+ {\rm sin}^2\alpha) .
\label{Eq11}
\end{eqnarray}
It is found that the force-free aligned rotator radiates with the Poynting flux of $L_{\rm aligned,FF}=\frac{\mu^2\,\Omega^4}{c^3}$. This contrasts with the vacuum aligned rotator that dose not radiate.

\begin{figure*}
\center
\begin{tabular}{cccccccccccccc}
\includegraphics[width=5.35 cm,height=6. cm]{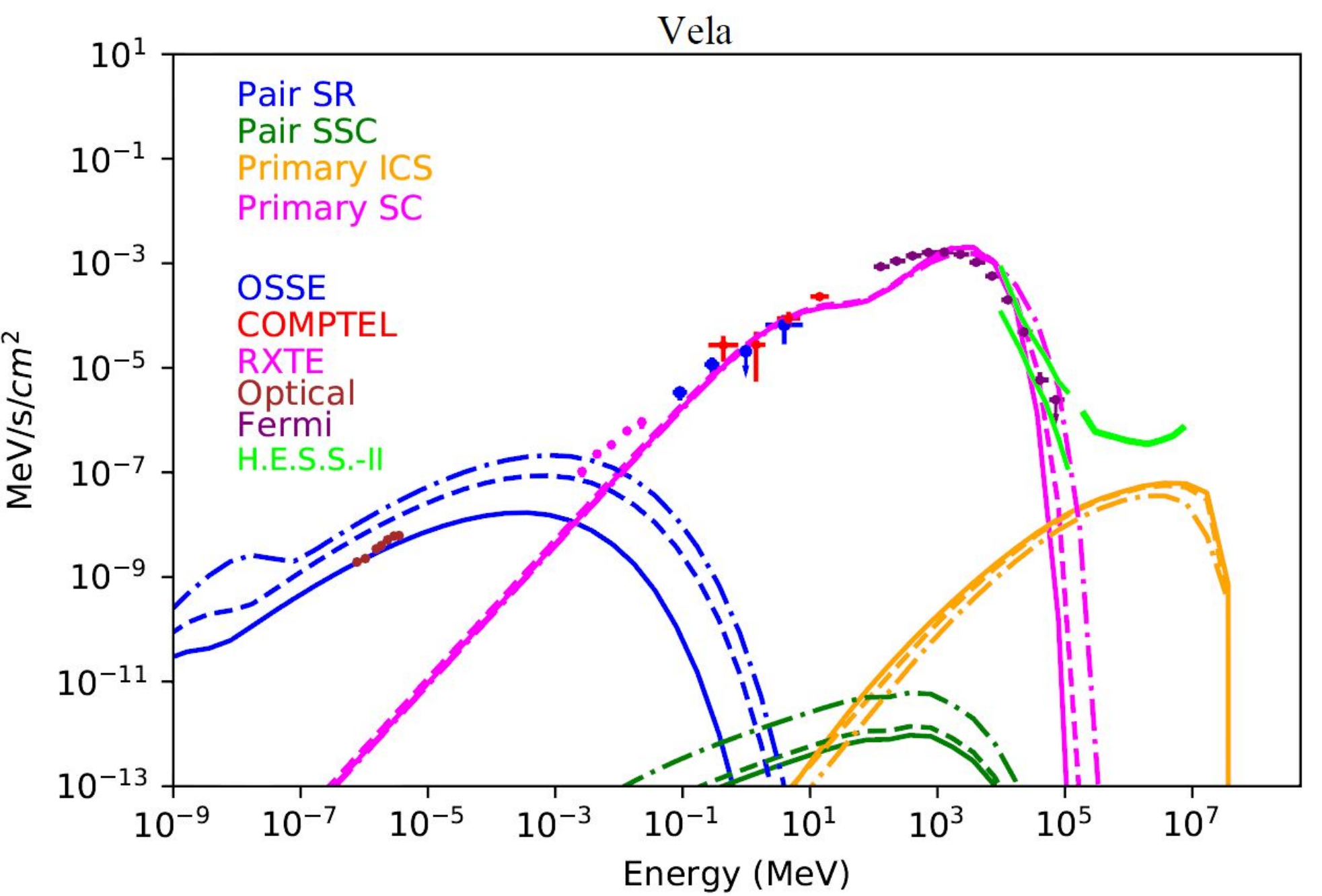} \,
\includegraphics[width=5.35 cm,height=6.05 cm]{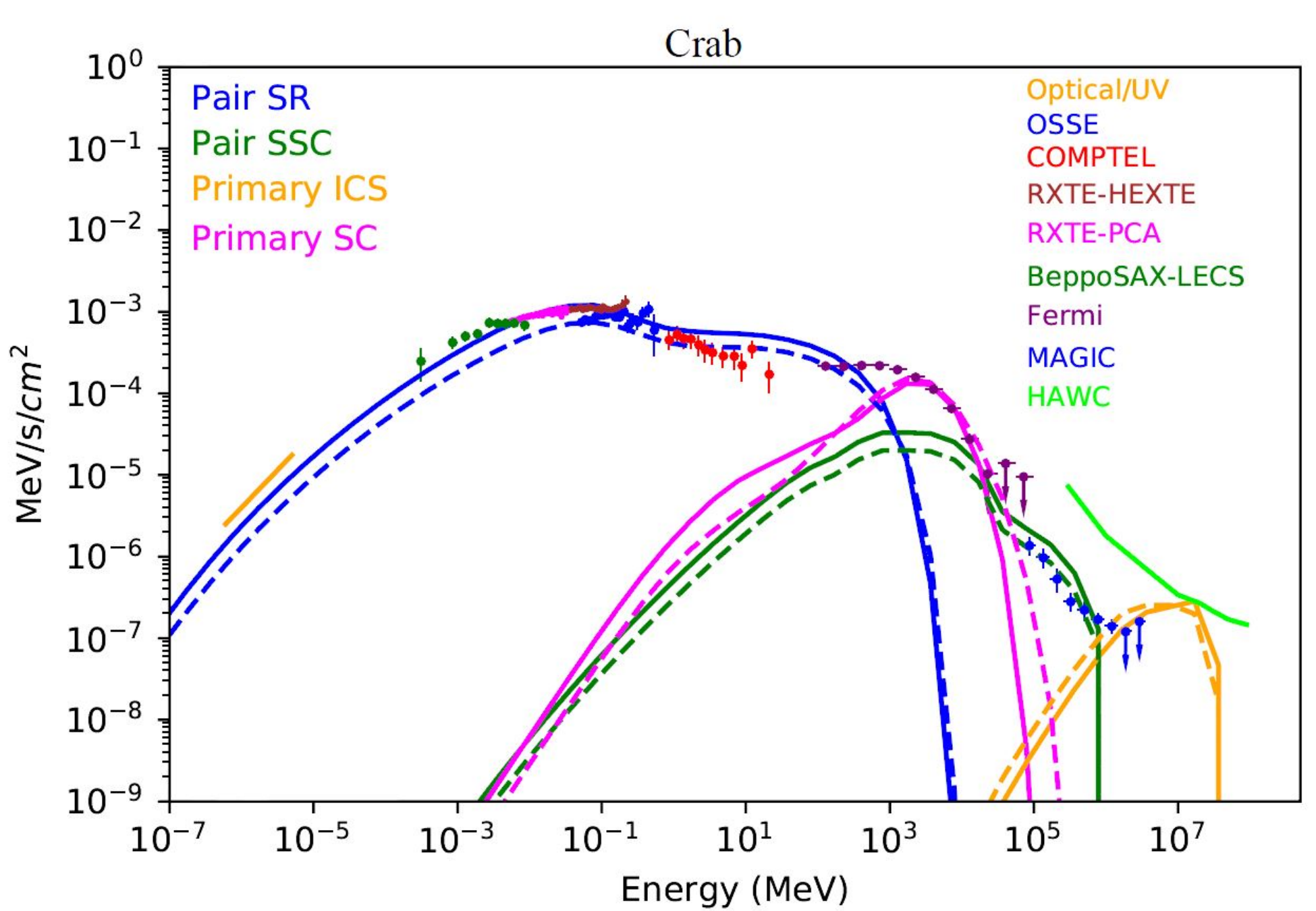} \,
\includegraphics[width=5.35 cm,height=6. cm]{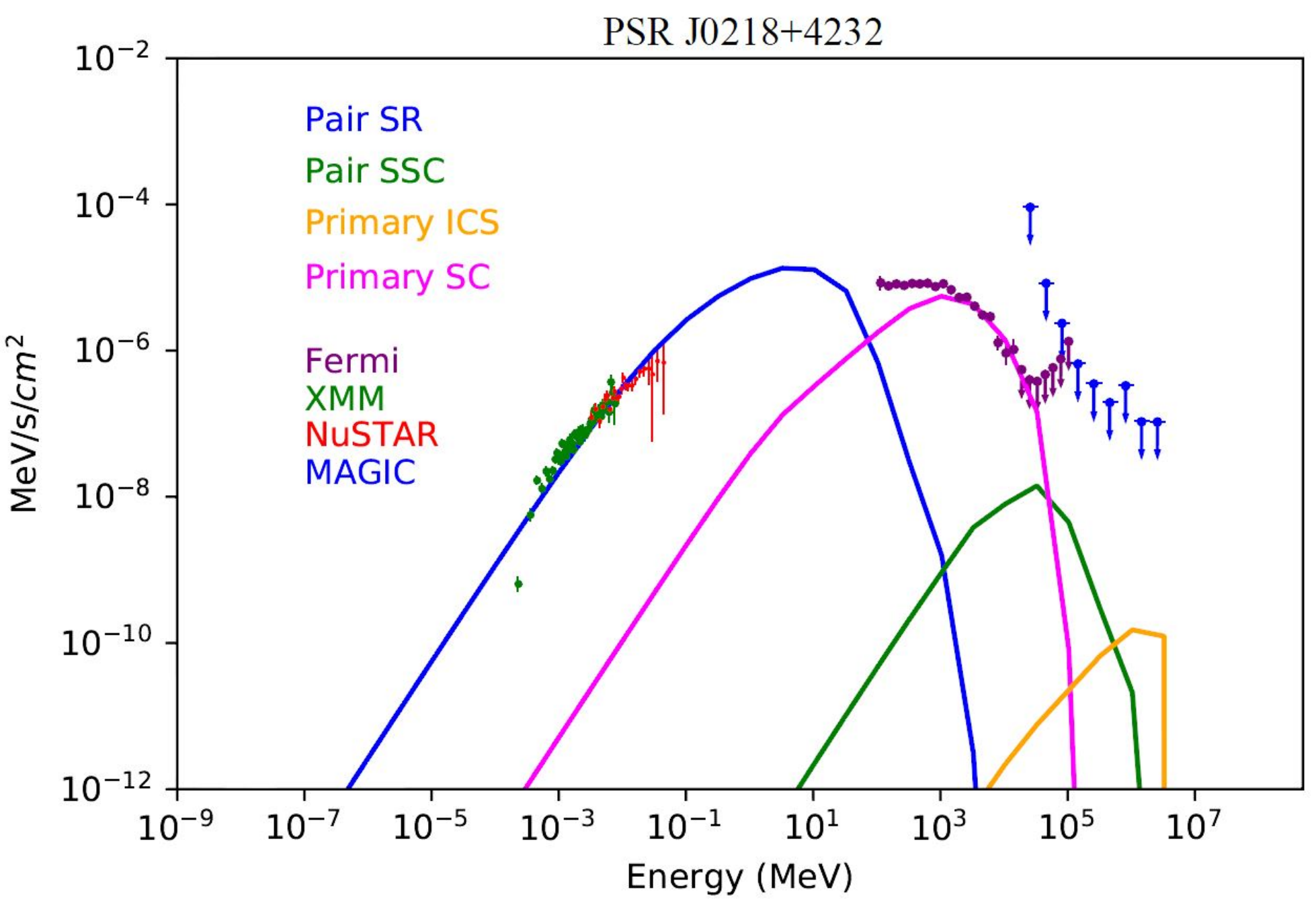} \,
\end{tabular}
\caption{The multi-wavelength emission modeling in the force-free magnetosphere for Vela with inclination angle $\alpha=75^{\circ}$ and viewing angles $\zeta=46^{\circ}$(solid), $46^{\circ}$(dashed), $70^{\circ}$(dotted–dashed) and the pair multiplicity $M_{+}=6\times10^3$, Crab with $\alpha=45^{\circ}$, $\zeta=60^{\circ}$(solid), $72^{\circ}$(dashed) and $M_{+}=3\times10^5$, and MSP PSR J0218+4232 with  $\alpha=60^{\circ}$, $\zeta=45^{\circ}$ and $M_{+}=3\times10^5$. The figure is taken from Harding, et al.  (2021) \citep{har21}. }
\label{fig8}
\end{figure*}

\subsection{Light curve and spectra modelling from the force-free magnetospheres}

The force-free solution provides more realistic field structures of pulsar magnetosphere than that of the vacuum solution, which can be used to model the pulsar light curves and spectra based on different radiation models with the assumed $\bm{E}_{\|}$ distributions. However, these models are not self-consistent since the force-free solutions do not allow any particle acceleration along magnetic field lines with $\bm{E}_{\|}=0$.
The force-free field structures was used to model the geometric $\gamma$-ray light curves by assuming the uniform emissivity along the particle trajectory in the observer frame \citep{bai10b}. The particle trajectory is defined as the combination of a drift velocity and one along the field line, which can help us compute the pulsar emission from the stellar surface to beyond the light cylinder. The projections of particle trajectories on the sky map were computed and  the geometric $\gamma$-ray light curves were produced based on the `separatrix' emission model extending from the stellar surface to beyond the LC \cite{bai10b}. The "separatrix" model forms the sky map stagnation (SMS) caustics, because the backward drift of the particle trajectory along asymptotically monopole field near or outside the LC compensates for the aberration and time delay effects. The emission from the same field line near the current sheet will arrive in phase to produce the caustics. It is shown that the "separatrix"  model can naturally produce the double-peak  $\gamma$-ray light curve in agreement with the Fermi observed ones. The  "separatrix" model with  the opposite hemisphere emission is similar to the SG model and is different from the one hemisphere OG model.
The geometric $\gamma$-ray light curves of the SG and OG model  were produced in the force-free and vacuum magnetosphere \citep{har11}. It is found that the peak phases of the force-free light curves are shifted to the later phases with the magnetic poles compared to the vacuum ones, which is caused by a larger offset toward the trailing side of the PC due to the larger PC size and higher field line sweepback (see Figure \ref{fig7}). The force-free light curves have  larger phase lags than those observed by Fermi-LAT, indicating that either the field structure is not close to the force-free one or the emission along the field lines is non-uniform. The force-free field was used to produce the radio and $\gamma$-ray light curves,  the magnetic inclination angles and viewing angles are then constrained through simultaneously fitting time-aligned radio and  $\gamma$-ray light curves \citep{ben21,pet21}. It is also shown that additional constraints can be obtained by fitting the radio polarization data with rotating vector model.

\begin{figure*}
\center
\begin{tabular}{cccccccccccccc}
\includegraphics[width=7.2 cm,height=7. cm]{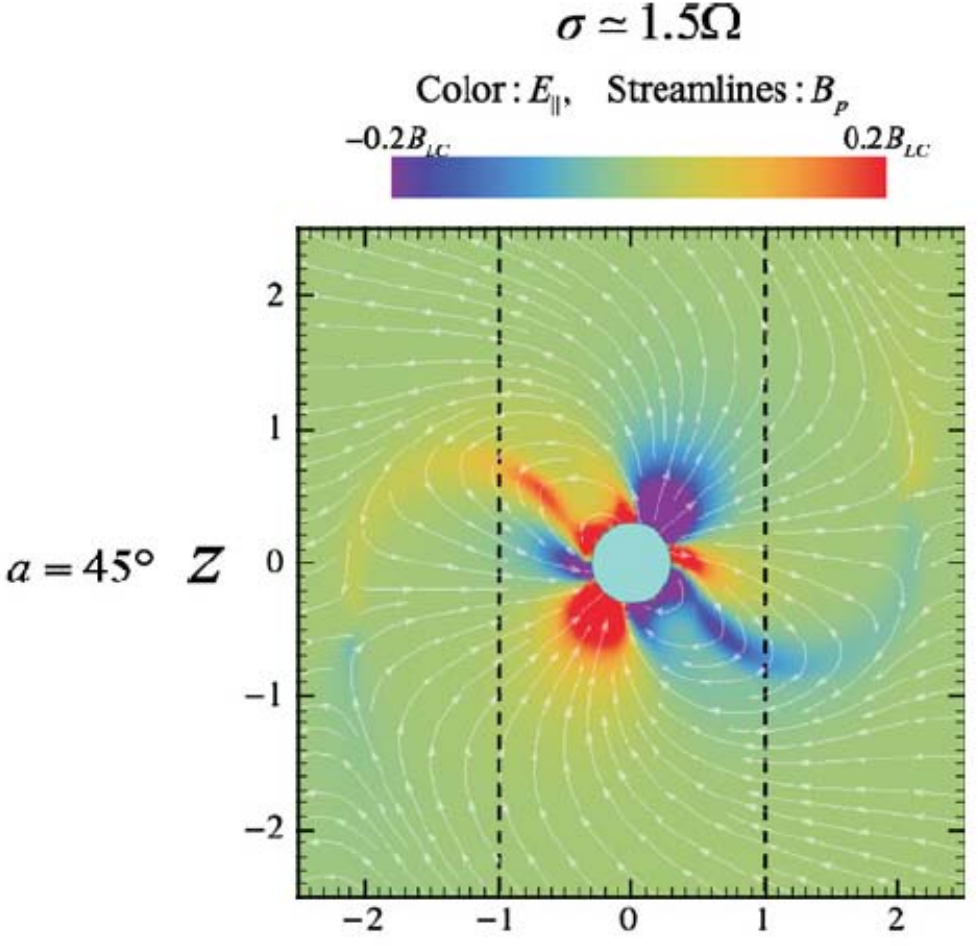} \qquad
\includegraphics[width=5.85 cm,height=7.05 cm]{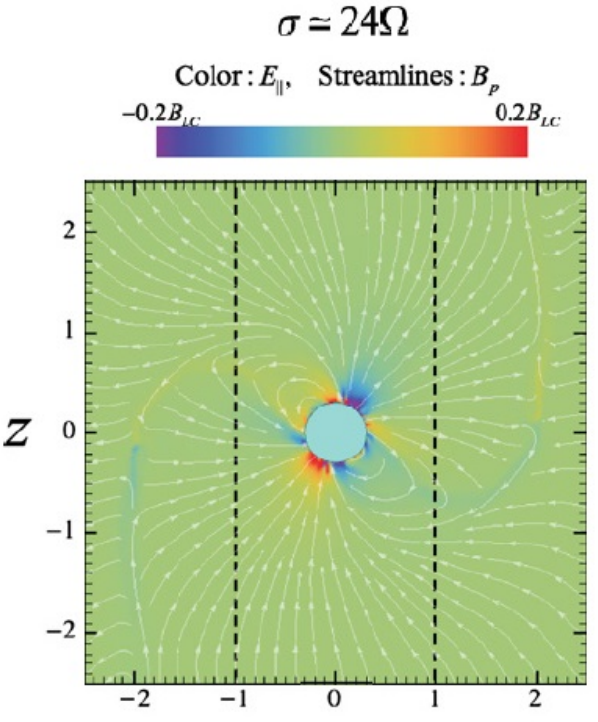}
\end{tabular}
\caption{The distributions of the magnetic field line and accelerating electric field for a $45^{\circ}$ resistive magnetosphere with $\sigma=1.5\,\Omega$ and $\sigma=24\,\Omega$ at the x$-$z plane. The figure is taken from Kalapotharakos et al. (2012a) \citep{kal12a}.}
\label{fig9}
\end{figure*}

The phase-resolved spectra and energy-dependent light curves of Vela pulsar were produced in an extended slot gap and current-sheet model for the force-free magnetosphere \citep{bar21}. The decreasing flux of the first $\gamma$-ray peak relative to the second one with increasing energy is attributed to the systematically larger curvature radius of the second peak than the first peak one. However, an additional phase shift needs to be added to match the peaks phases of the observed $\gamma$-ray light curves. Therefore, an  azimuthal  dependent $\bm{E}_{\|}$ distribution are suggested to give an improved match over the observed $\gamma$-ray light curves \citep{bar21}. The particle trajectory method was used to model the multi-wavelength emission from the young pulsars and MSPs in the force-free magnetosphere \cite{har15,har21}  (see Figure \ref{fig8}), where the synchrotron, curvature, synchrotron-self Compton (SSC) emission from both the primary particles accelerated with an assumed $\bm{E}_{\|}$ value near the current sheet and electron–positron pairs produced in the PC cascades were computed.
The synchrotron emission from the  pairs produces the optical to hard X-ray spectrum. The SSC emission from the  pairs  produces a broad spectrum with typically peak energy between 1 and 10 GeV for most young pulsars and around 100 GeV for MSPs, and extends to around 1 TeV for Crab.
The pair SSC emission from MSPs peaks at the higher energy than most young pulsars due to higher pair energy, but the MSPs SSC component is suppressed to the low flux level by the Klein-Nishina effects. Therefore, the SSC emission from the pairs have a negligible contribution to the observed spectrum for most young pulsars and  MSPs, only the  Crab-like pulsars  produce the detectable SSC components. The curvature emission from the primary particles contributes to  the Fermi spectrum from several GeV up to 100 GeV and the tail of this emission can explain the  HESS-II and MAGIC spectrum  from Vela, Geminga and B1706-44. The SSC emission from the primary particles produce a very high-energy spectrum up to around 30 TeV \citep{har18}. This spectrum component is well below current detection thresholds for most pulsars, but it is still detectable for the Crab and other pulsars by the  High Altitude Water Cherenkov Observatory and Cherenkov Telescope Array.
The detected maximum photon energy can provide a direct lower limit on the maximum energy of the accelerated particles. The maximum photon energies above 1 TeV from Crab and Vela requires the very high particle energies of $\gamma>10^7$, which tends to favor the curvature radiation over the synchrotron radiation as the dominant Fermi $\gamma$-ray emission mechanism. The pulsar polarization measurements can provide an independent constraint on the location and geometry of the emission region and emission mechanism. The pulsar multi-wavelength polarization was also modelled in the force-free field structure \citep{har17}, including synchrotron emission at the optical to hard X-ray band from the  cascade pairs  and the synchrotron or curvature emission in the $\gamma$-ray band from the primary particles. It is found that a fast position angle (PA) swing and depolarization is produced during the light-curve peaks in all energy bands. The PA swings and  depolarization increase as the emission radius increase, there is a  nearly $180^\circ$ PA swing for emission in the current sheet outside the light cylinder due to the change of the magnetic polarity. The curvature emission is predicted to produce a higher polarization degree than the synchrotron or inverse-Compton emission. Therefore, the $\gamma$-ray polarization measurements can  distinguish whether the synchrotron or curvature emission is the dominant Fermi $\gamma$-ray emission mechanism. It is also suggested that detection of a sudden change in PA and a sharp rise in polarization degree at 1-100 MeV may indicate that curvature emission is the Fermi $\gamma$-ray  emission mechanism.

\section{The Resistive magnetospheres}
\label{sec6}
\subsection{Field structure of the resistive magnetospheres}

The force-free solutions provide a useful first step to understand more realistic plasma-filled magnetosphere. However, the force-free solutions cannot allow any accelerating electric fields along magnetic field lines, they cannot provide any information about the  particle acceleration and the production of radiation. A more realistic pulsar magnetosphere should allow for a local dissipative region in the magnetosphere to produce the observed pulsed emission. A new degree of freedom is required to include the dissipation effect in the magnetosphere. The vacuum solutions have $\bm{E}_{\|}$  but has no particles to be accelerated, while the force-free solutions have  enough particle  but no $\bm{E}_{\|}$ to accelerate the particles. It is motivated that the realistic pulsar magnetosphere should lie between the vacuum limit and force-free limit. Therefore, the resistive magnetospheres are developed to include the dissipation effect by introducing a conductivity parameter, which can span the magnetospheric solutions from the vacuum to force-free limit and the spatial distribution of the $\bm{E}_{\|}$ value is  self-consistently controlled by the  $\sigma$ value. The current density can be defined  by the conductivity parameter $\sigma$ in an Ohm's law form. However, there is no unique prescription for the exact expression of the resistive current density. A formulations of the resistive current density can be defined by the Ohm's law in the fluid rest frame where the $\bm{E}$ and $\bm{B}$ fields are parallel, the current density $\bm{J}$ in the observer frame can be obtained by the Lorentz transform with the minimal velocity hypothesis \citep{li12}
\begin{eqnarray}
{\bm J}= \frac{ \rho_{\rm e} {\bm E} \times {\bm B} + \sqrt{\frac{B^2+E^2_0}{B^2_0+E^2_0}}\sigma E_0(B_0{\bm B}+E_0{\bm E)}} { B^2+E^2_{0}},
\label{Eq12}
\end{eqnarray}
where $E_0$ and $B_0$ are defined  by the Lorentz invariant relations
\begin{eqnarray}
B^2_{0}-E^2_{0}={\bm B}^2-{\bm E}^2,  \, E_{0}B_{0}={\bm E}\cdot {\bm B}, \, E_{0}\geq0.
\label{Eq13}
\end{eqnarray}
A simpler current density  is also defined in the Ohm's law in the observer frame by \citep{kal12a,cao16b}
\begin{eqnarray}
{\bm J}=  \rho_{\rm e} {{\bm E} \times {\bm B} \over B^2+E^2_{0}}+\sigma {\bm{E}}_{\|}\;,
\label{Eq14}
\end{eqnarray}
It is expected that different current density description should obtain similar field line structures and $\bm{E}_{\|}$ distributions.

\begin{figure*}
\center
\begin{tabular}{cccccccccccccc}
\includegraphics[width=16. cm,height=10. cm]{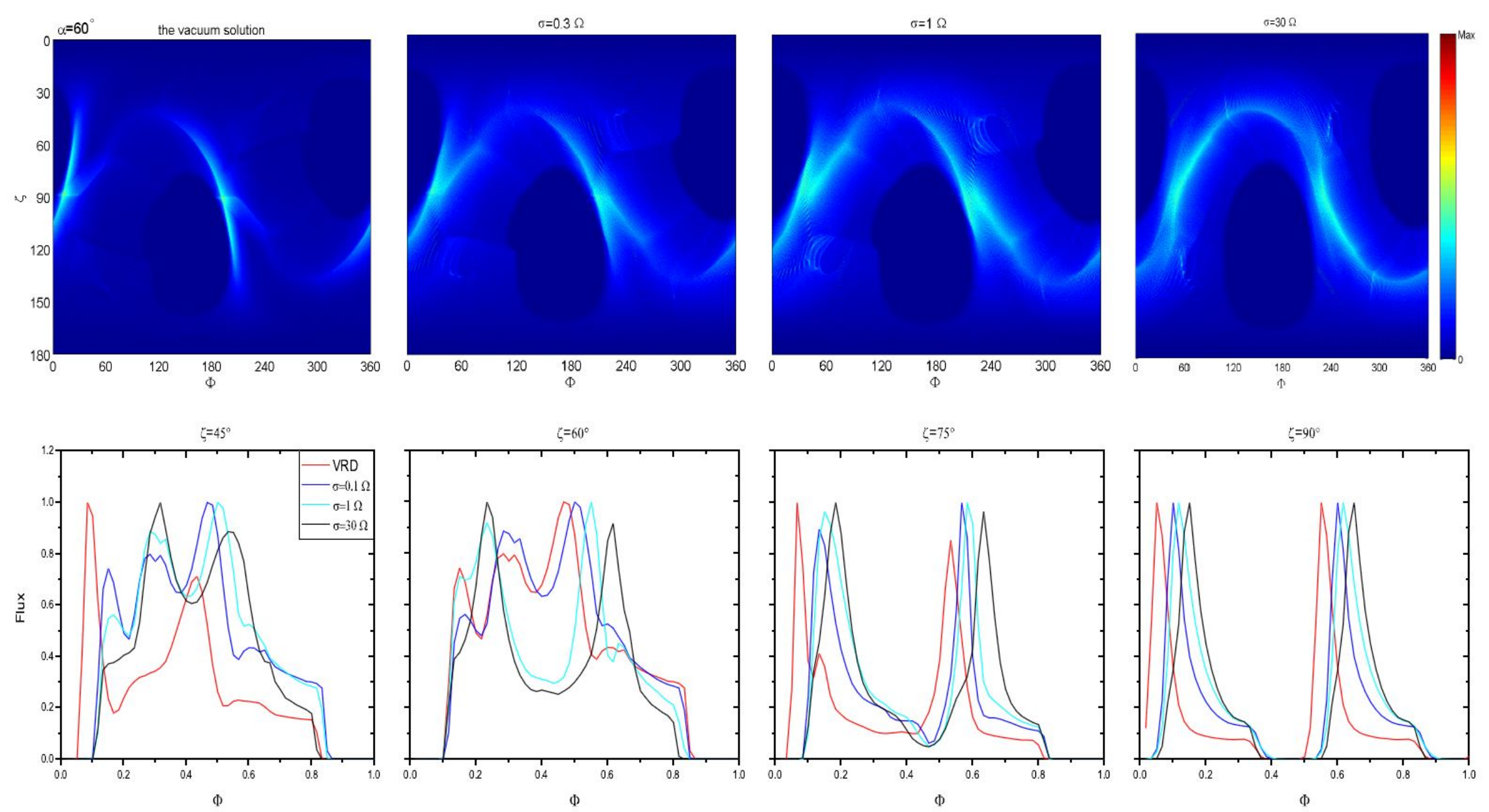}
\end{tabular}
\caption{The geometric sky maps and the corresponding $\gamma$-ray light curves for a $60^{\circ}$ inclined rotator at different view angles in different  resistive magnetospheres. The figure is taken from Cao \& Yang  (2019) \citep{cao19}.}
\label{fig10}
\end{figure*}

The time-dependent Maxwell equation (\ref{Eq1}) with the resistive current density (\ref{Eq12}) and (\ref{Eq14}) were numerically solved by  the FDTD method \citep{li12,kal12a} and  followed by the spectral method \citep{cao16b}. All these time-dependent simulations obtain a range of similar resistive solutions, which spans  from the vacuum solution to the force-free solution with increasing $\sigma$ and show the self-consistent $\bm{E}_{\|}$ distributions in the magnetospheres. Figure \ref{fig9} shows the distributions of the magnetic field line and accelerating electric field for a $45^{\circ}$ resistive magnetosphere with $\sigma=1.5\,\Omega$ and $\sigma=24\,\Omega$ at the x$-$z plane. It is found that the field line structure is close to the vacuum solution and a strong $\bm{E}_{\|}$ region  appears only inside the LC for low  $\sigma$. As the conductivity $\sigma$ increase, the resistive solution  tends to the force-free solution with increasing PC size and field line sweepback, which shift the PC towards the trailing side. A strong $\bm{E}_{\|}$ region also appears near the Y-shaped current sheet outside the LC. The $\bm{E}_{\|}$  decreases with increasing  $\sigma$ and the $\bm{E}_{\|}$ region only exist in the PC region inside the LC and  near the Y-shaped current sheet outside the LC for the high $\sigma$. Moreover, the Poynting flux also increases with increasing conductivity $\sigma$ and inclination angle $\alpha$.

\subsection{Light curve and spectra modelling from the resistive magnetospheres}

\begin{figure*}
\center
\begin{tabular}{cccccccccccccc}
\includegraphics[width=6.1 cm,height=6.3 cm]{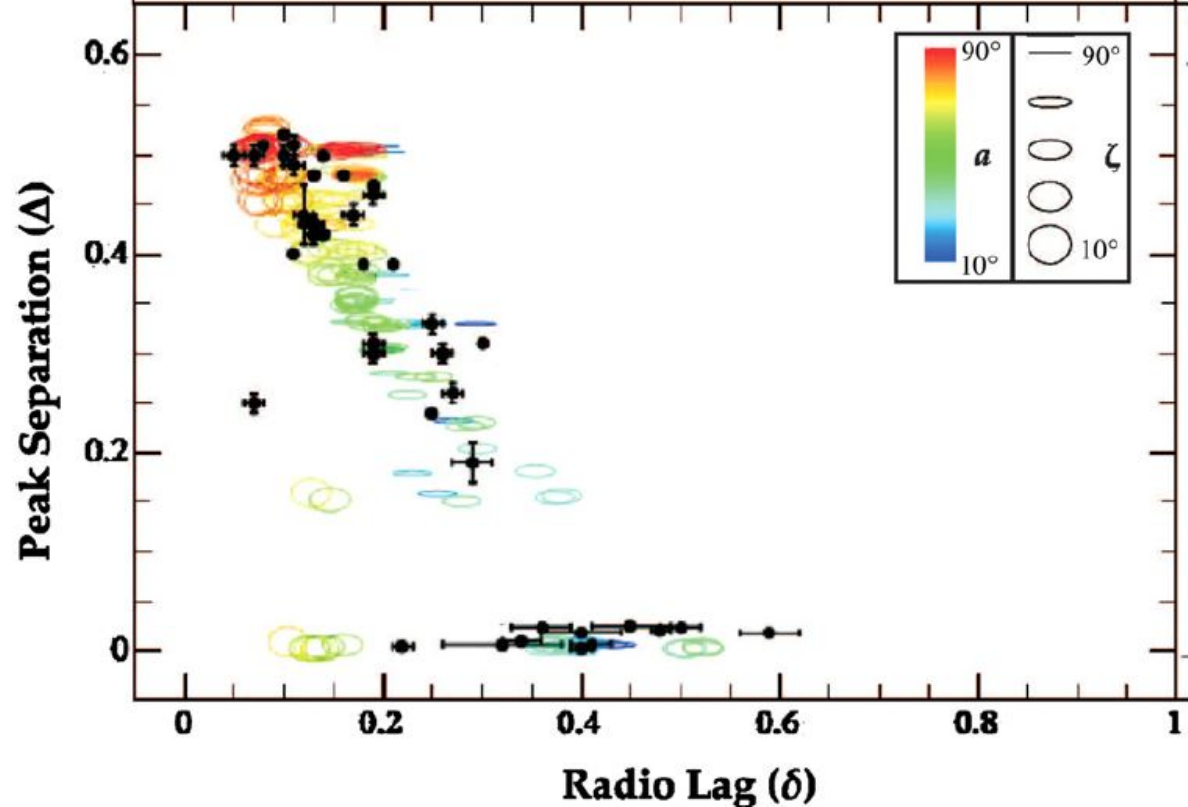} \qquad
\includegraphics[width=6.5 cm,height=6.5 cm]{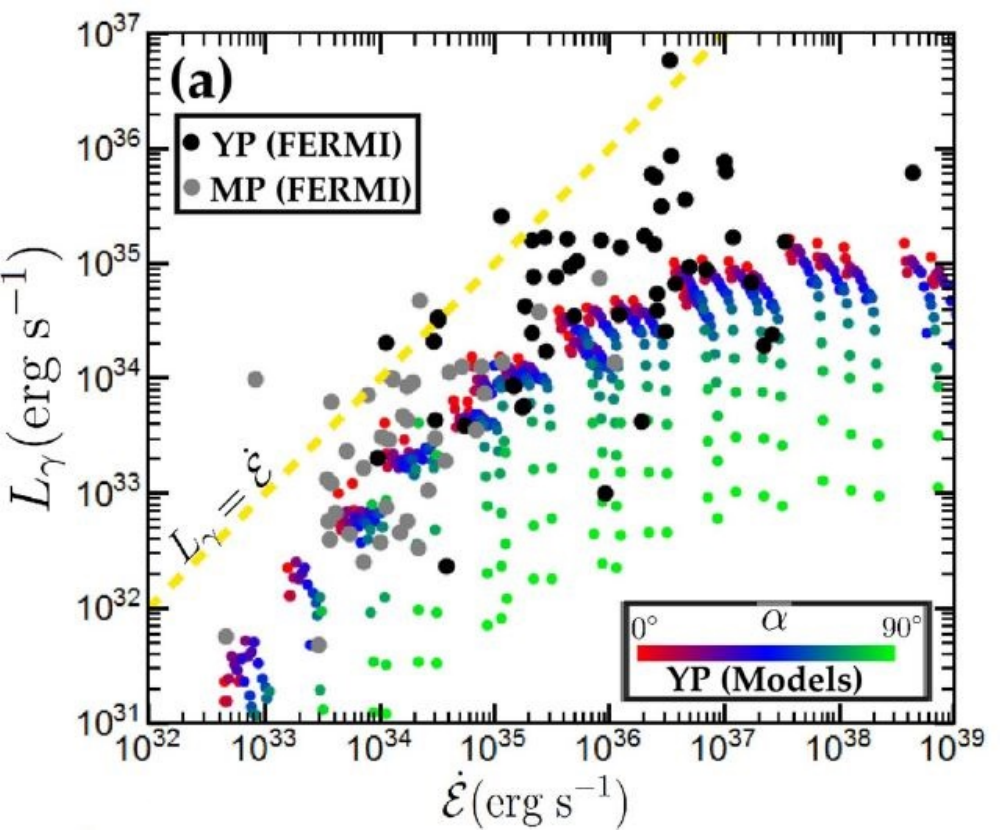}
\end{tabular}
\caption{The predicted $\delta-\Delta$ and $L_{\gamma}-\ed$ distribution from the FIDO model. The figure is taken from Kalapotharakos et al. (2014) \citep{kal14} and Kalapotharakos et al. (2017) \citep{kal17}. }
\label{fig11}
\end{figure*}

The resistive magnetospheres not only provide the field structure geometries but also the self-consistent $\bm{E}_{\|}$ distributions from the vacuum  to the force-free solutions, which can be used to model the pulsar $\gamma$-ray light curves and spectra by the direct comparison with observations. By assuming the uniform emissivity along the field lines in the comoving frame, the resistive magnetospheres were used to model the geometry $\gamma$-ray light curves \citep{kal12b}. The geometry $\gamma$-ray  light curves of the OG and SG model were produced for a range of the resistive magnetospheres  from the vacuum  to the force-free solution. It is found that the  peak phase of the $\gamma$-ray  light curves are progressively shifted to the later phase  with respect to the magnetic pole as the $\sigma$ increases, which is attributed to  a larger offset toward the trailing side of the PC with increasing  $\sigma$. The geometry $\gamma$-ray  light curves were also produced from the spectral method simulation in the resistive magnetospheres \citep{cao19}, a similar progression in the $\gamma$-ray  light curve shape as  $\sigma$ increases is also found (see Figure \ref{fig10}). Based on the self-consistent  $\bm{E}_{\|}$ distributions from the resistive magnetospheres, the $\gamma$-ray  light curves were first modeled through computing the curvature radiation along the particle trajectory \citep{kal12b}. It is found that these light curves generally have a larger phase lags than those in the geometry model. A further exploration of the $\gamma$-ray emission pattern and light curves from the curvature radiation in resistive pulsar magnetospheres was performed by \citep{kal14}, it is found that all the emission comes from the inner magnetosphere for the low $\sigma$ values. As the $\sigma$ increases, the emission gradually moves outward to the high attitude, and a significant part of the emission is produced in regions near the Y-shape current sheet outside the LC.  All the emission is produced in regions near the Y-shape current sheet outside the LC for the very high $\sigma$ values. The light curves are broad for the low $\sigma$ values and significantly become narrower as  the $\sigma$ increases. By performing the  comparison of the  model radio-lag ($\delta$) versus peak-separation ($\Delta$) distribution and the observed one, it was found that the low $\sigma$ models give the poorest match to the observed  $\delta-\Delta$ distribution \citep{kal14}.  A relatively good match over the observed $\delta-\Delta$ distribution can be obtained for the high $\sigma$ models with the emission from the current sheet outside the LC , but there are still some model points on the $\delta-\Delta$  diagram that do not lie near the observed ones. Motivated by the  results of the high $\sigma$ models, the Force-free Inside Dissipative Outside (FIDO) model with the near-force regime inside the LC and the dissipative regime outside the LC is developed to produce the $\gamma$-ray  light curves. It is found that the FIDO model can best explain  the light curve characteristics of the Fermi $\gamma$-ray pulsars, particularly the observed $\delta-\Delta$ distribution (see Figure \ref{fig11}). The FIDO model shows the emission from the 3D current sheet is non-uniform and non-axisymmetric.  The self-consistent $\bm{E}_{\|}$ distribution from the resistive magnetosphere was used to model the $\gamma$-ray  light curves from the curvature radiation \citep{cao19}, it is found that the FIDO model with the non-uniform emission in the current sheet provides a better match to the observed $\gamma$-ray light curves.

In the frame of the FIDO model with a range of conductivity $\sigma$ outside the LC, the properties of the energy-dependent $\gamma$-ray light curves and spectra from Fermi pulsars were explored by \citep{bra15}. It was found that the variation of the spectral index and cutoff energy as a function of pulse phase can be explained reasonably. The FIDO model can also reproduce the observed trend of the $\gamma$-ray light curves with  the  decreasing ratio of the first to second $\gamma$-ray peak toward higher energies. It is indicated that the conductivity $\sigma$ increases with the spin-down power $\ed$, which is expected that the  higher $\ed$ pulsars have more efficient pair cascades that can provide the magnetosphere conductivity \citep{bra15}. However, this study only uses an approximate $\bm{E}_{\|}$ values from the corresponding force-free solutions instead of the resistive solutions. To restrict the $\bm{E}_{\|}$ regions  only near the current sheet outside the LC, the FIDO model was refined through applying magnetic-field-line dependent conductivity $\sigma$ \citep{kal17}.
A relation between  $\sigma$ and  $\ed$ was found  by reproducing the observed  cutoff energy with the refined FIDO model. It is shown that the $\sigma$ increases with $\ed$ for the high $\ed$ and saturates for the low $\ed$. The refined FIDO model  can  reproduce the observed trends of the  $\gamma$-ray luminosity $L_{\gamma}$ as a function of $\ed$. It is suggested that a higher emitting particle multiplicities are needed to reproduce the observed $L_{\gamma}$ at the high $\ed$ (see Figure \ref{fig11}). Moreover, the FIDO model with the self-consistent $\bm{E}_{\|}$ values from the resistive magnetosphere was used to explore the energy-dependent $\gamma$-ray light curves and spectra from the Fermi pulsars \citep{yan21}. It is  also found that the emission from the FIDO model is non-uniform and asymmetric over the PC and the high emission is concentrated on the leading side of the PC edges, and the FIDO model  can reproduce the phase-averaged spectra and energy-dependent $\gamma$-ray light curves for the Vela and Crab pulsar, especially the observed trend of the decreasing ratio of the first to second $\gamma$-ray peak with increasing energies \citep{yan21}.

\begin{figure*}
\center
\begin{tabular}{cccccccccccccc}
\includegraphics[width=10.5 cm,height=10.5 cm]{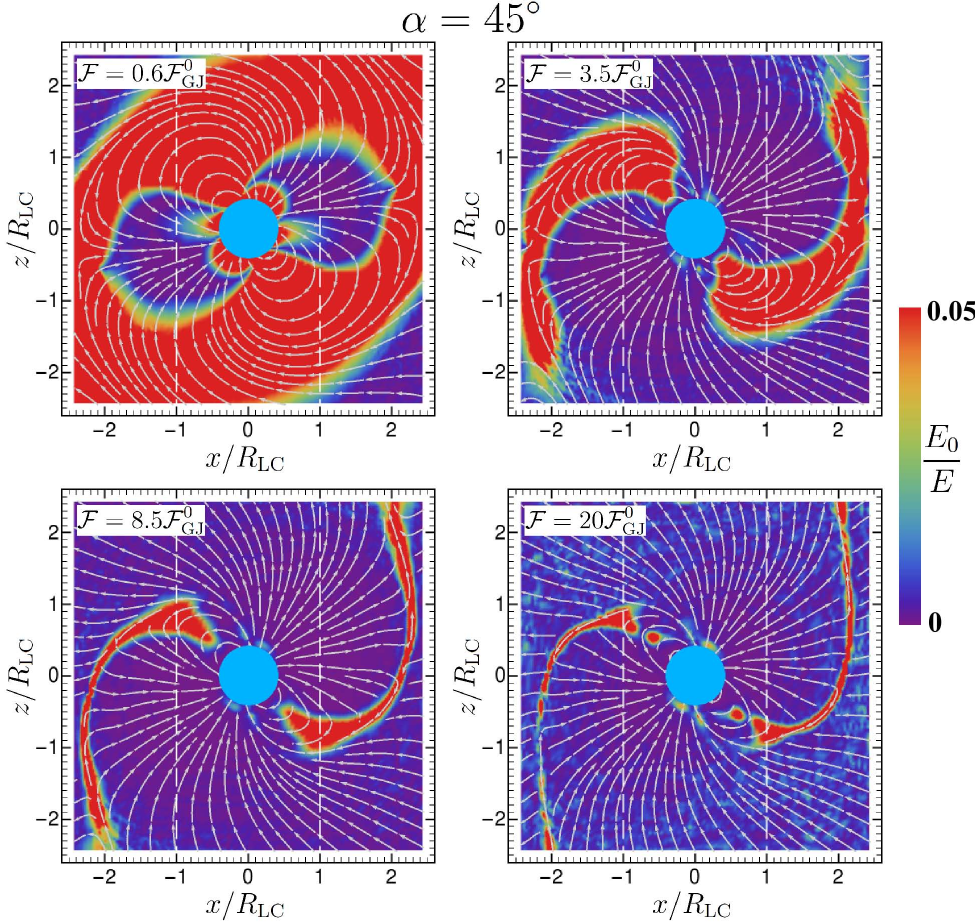}
\end{tabular}
\caption{The distributions of the magnetic field line and accelerating electric field in the PIC simulation of a $45^{\circ}$ inclined rotator with different particle injection rates at the x$-$z plane. The figure is taken from Kalapotharakos et al. (2018) \citep{kal18}}
\label{fig12}
\end{figure*}

\section{The PIC magnetospheres}
\label{sec7}

\subsection{Field structure of the PIC magnetosphere}

The resistive magnetospheres can produce the accelerating electric field that are self-consistent with the magnetic field structures by introducing a macroscopic conductivity. However, the resistive magnetospheres are still not self-consistent since they can not model the source of particles that provide magnetosphere conductivity. The kinetic plasma simulation is required to self-consistently treat the  feedback between the particle motions, radiating photons, and electromagnetic fields.  The kinetic model provides a method to self-consistently compute the feedback between the particle motions, radiating photons and electromagnetic fields by solving the time-dependent Vlasov-Maxwell equations. The time-dependent Vlasov equations are given by
\begin{eqnarray}
\frac{\partial f }{\partial t}+\frac{\bm{p}}{\gamma\,m}\frac{\partial f }{\partial \bm{r}}+q\left(\bm{E}+\frac{\bm{v}\times\bm{B}}{c}\right)\,\frac{\partial f }{\partial \bm{r}}=0,
\label{Eq15}
\end{eqnarray}
where $\bm{p}$ is the particle momentum and $f(\bm{r},\bm{p})$ is particle distribution function.
It is difficult to directly solve the time-dependent Vlasov-Maxwell equations  in  the six-dimensional phase space $f(\bm{r},\bm{p})$. A particle-in-cell (PIC) method can be used to indirectly solve the time-dependent Vlasov-Maxwell equations by sampling  the distribution function in phase space with the particles. Each particle represents a large number of real particles that approximately following the exact same path in phase space. The Vlasov equation can be solved along characteristics curves by a set of ordinary differential equations
\begin{equation}
\begin{split}
\frac{d \bm{r} }{d t}&=\bm{v}\;,\\
\frac{d \bm{v} }{d t}&=q\left(\bm{E}+\frac{\bm{v}\times\bm{B}}{c}\right)\;.
\end{split}
\label{Eq16}
\end{equation}
which corresponds to the particle motion equations. The particle motion equations (\ref{Eq13}) produce an set of characteristic curves  representing a surface in phase space, which is a solution of the Vlasov equation. The currents and charges from the particle motions are deposited in the numerical grid by following the charge-conserving scheme
\begin{equation}
\begin{split}
\rho&=\sum_{j}q_j\delta(\bm{r}-\bm{r}_j)\;,\\
\bm{J}&=\sum_{j}q_j\bm{v}_j\delta(\bm{r}-\bm{r}_j)\;.
\end{split}
\label{Eq17}
\end{equation}
which can be used as the sources to solve the Maxwell equations.

The time-dependent Maxwell equations (\ref{Eq1}) and the motion eqautions (\ref{Eq17}) for an aligned rotator was first solved  through the 2D Cartesian PIC Tristan-MP code by Philippov \& Spitkovsky (2014) \citep{phi14}, the structure of pulsar magnetospheres was explored  by injecting the particles only from the stellar surface or everywhere in the magnetosphere.
The PIC simulation of an aligned rotator was performed through the 2D spherical PIC code with the incorporation of the full pair cascades from the stellar surface to beyond the LC \citep{che14}. The  PIC simulations of  an aligned rotator by  the 2D spherical PIC code were also performed with the particle injection from the $\bm{E}_{\|}\neq0$ region \citep{che14} or only above the stellar surface \citep{cer15}. Further, the PIC simulations of an oblique rotator were first performed through the 3D Cartesian PIC Tristan-MP code by Philippov \& Spitkovsky (2015) \citep{phi15a}. In this case, an approximate pair cascades by injecting the particles whose energy exceeds a given threshold is implemented.
The general-relativistic corrections with the frame-dragging and time–space curvature were later included  in the 3D Cartesian PIC Tristan-MP code \citep{phi15b,phi18}.
The  PIC simulations of an oblique rotator were also performed by the 3D spherical PIC Zeltron code with the particle injection from only above the stellar surface \citep{cer16a}. These PIC simulations show that magnetospheres transition from a charge-separated "electrosphere" solution with the disk–dome structure for the low  particle injection to a near force-free solution with the particle acceleration near the  current sheet  for the high particle injection. It is also found that the pair cascades from the stellar surface can not fill the magnetosphere to produce a near force-free solution, the pair cascades in the current sheet is needed to produce a near force-free solution. Recently, a 3D Cartesian PIC code was developed to perform the PIC simulations of an oblique rotator \citep{kal18}, and the magnetosphere transition from the near vacuum solution to the near force-free solution was explored by applying the particle injection rates from the low to high values. Figure \ref{fig12} shows the distributions of the magnetic field line and accelerating electric field in the PIC simulation of a $45^{\circ}$ inclined rotator with different particle injection rates at the x$-$z plane. It is shown that  the $\bm{E}_{\|}$ regions are confined only near the  current sheet outside the LC as the particle injection rate increases, while the volume of the $\bm{E}_{\|}$ region decreases with increasing  particle injection rate. A similar trend of the $\bm{E}_{\|}$ distribution with increasing particle injection rate was also found by applying the magnetic-field-line dependent particle injection in the PIC simulation \citep{kal23}. It is also found that a near force-free magnetosphere can be produced at all inclination angles without the need for pair cascades near the current sheet \citep{bra18}. All the PIC simulations show a near force-free magnetosphere with the particle acceleration near the current sheet, revealing that the particle acceleration and the high-energy emission is mainly produced near the current sheet outside the LC. Moreover,  the reported  Poynting flux from the PIC simulation is  similar to the MHD expectation.  All the PIC simulations show the $\sim$$1-20\%$ dissipation of the Poynting flux outside the LC, which  is converted into particle acceleration and radiation in the current sheet.

\begin{figure*}
\center
\begin{tabular}{cccccccccccccc}
\includegraphics[width=8. cm,height=8. cm]{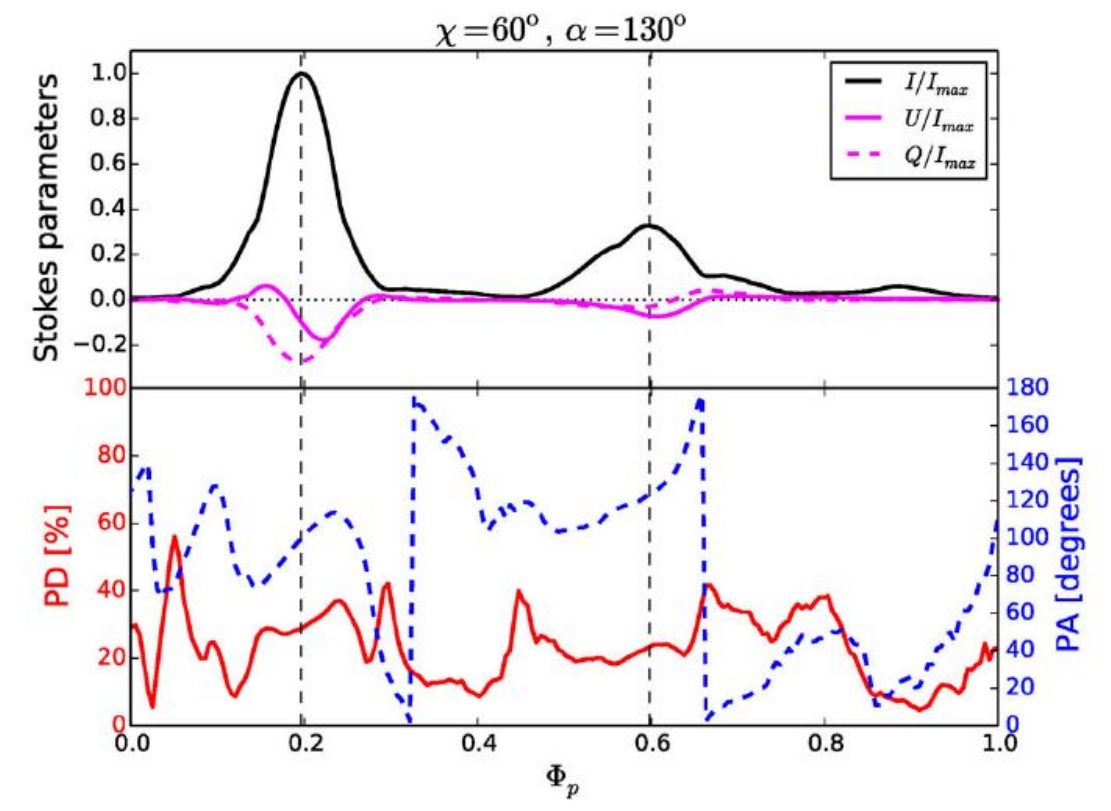}
\end{tabular}
\caption{The predicted synchrotron $\gamma$-ray polarization properties from the PIC simulation for the Crab-like pulsar. The figure is taken from Cerutti et al. (2016b) \citep{cer16b}.}
\label{fig13}
\end{figure*}

\subsection{Light curve and spectra modelling from the PIC magnetospheres}

The PIC  magnetospheres have the near force-free field structures with the $\bm{E}_{\|}$ regions near the current sheet outside the LC, which can be connected to the pulsar observational signatures by simultaneously extracting the pulsar emission information from the current sheet in the PIC simulation. The PIC simulation can not use the realistic physical  parameters to model the pulsar magnetosphere because of the large scale separation between  the plasma frequency and the pulsar rotation frequency. The artificially low magnetic field of $B\sim10^5 \, \rm G$ is  usually used in the PIC simulations to resolve the plasma frequency, so that the particles are only accelerated  to the low energy of $\gamma\sim10^3$. This is clearly not enough to produce the observed $\gamma$-ray photons  by the synchrotron or curvature radiation. Therefore, a scale-up technique for the magnetic field and energy is required to model the pulsar $\gamma$-ray emission. The pulsar $\gamma$-ray emission was studied by  modelling the synchrotron radiation from the PIC particles accelerated by magnetic reconnection in the current sheet \citep{cer16a,phi18}, which can produce the $\gamma$-ray light curves and spectra by scaling up the synchrotron photons of the PIC particles to those of the real pulsar. It is found that the sky maps from the PIC model show the two bright caustics that are characteristic of radiation from the current sheet. The double-peak $\gamma$-ray light curves can generally be produced when the observer angle crosses the two bright caustics, but the phase lags from the magnetic pole are seemly larger that those seen in Fermi $\gamma$-ray light curves. The predicted $\gamma$-ray spectra show the power-law exponential cutoff shapes with the cut-off energies around several GeV, which are similar to those of the Fermi $\gamma$-ray spectra. The pulsar $\gamma$-ray emission was also explored by modelling the curvature radiation from the scaled-up PIC particles \citep{kal18,kal23}, where the surface magnetic field, the accelerating electric field, and the particle energy are scaling up to those of the real pulsar. It is found that the particles can be accelerated to the realistic emitting $\gamma$-ray energies of $\gamma\sim10^7$ in the radiation-reaction limited regime where the particle acceleration is balanced by the curvature losses, the predicted $\gamma$-ray spectra with the cutoff energies of several GeV  are also similar to those of the Fermi $\gamma$-ray spectra. Moreover, the predicted $\gamma$-ray light curves have smaller phase lags from the magnetic pole than those of the PIC synchrotron model, which is better agreement with those of the Fermi $\gamma$-ray light curves. It is also found that a positive relation between the the particle injection rate and the spin-down power by reproducing the observed $\gamma$-ray cutoff energies and luminosities, which provides a physical link between the microscopic injection rate and macroscopic conductivities. The pulsar $\gamma$-ray emission was further explored by the refined PIC model with the magnetic-field-line dependent particle injection, it is shown that the PIC magnetosphere with the particle injection near the separatrix zone can better reproduce the broad properties of the Fermi $\gamma$-ray pulsar phenomenology, including the $\gamma$-ray light curve profiles, the spectra shapes, and the $\delta-\Delta$ relation \citep{kal23}.

The pulsar $\gamma$-ray polarization was also predicted by modelling the synchrotron radiation from the current sheet in the PIC simulation  \citep{cer16b}. Figure \ref{fig13} shows the predicted $\gamma$-ray polarization properties from the PIC synchrotron model for the Crab-like pulsar. It is found that each pulse peak is accompanied by a nearly $180^\circ$ swing of the polarization angle. This is the natural result of the emission pattern from the current sheet where the the field line polarity changes as the line of sight crosses the current sheet. It is also shown that the synchrotron radiation predicts a lower  polarization degree ($\lesssim 30\%$) than that of the curvature curvature ($\gtrsim 40\%$). The future $\gamma$-ray polarization measurements can help us distinguish the  curvature origin from the synchrotron one for the observed Fermi $\gamma$-ray emission.

\section{The combined force-free and AE magnetospheres}
\label{sec8}

\subsection{Field structure of the combined force-free and AE magnetospheres}

The PIC simulation cannot use the real pulsar parameters to model the pulsar magnetosphere and predict the subsequent radiation because of the large ratio between the macroscopic scales to the microscopic scales. Therefore, the scale-up method by extrapolating the simulation parameters to real pulsar parameters is required to model the pulsar high-energy emission. However, it is dangerous for any extrapolation to the realistic values in the non-linear emission scenarios. Therefore, a clever method is required to perform realistic  simulation of the pulsar magnetosphere  and predict the subsequent pulsed emission.

It is expected that the pulsar magnetosphere is filled with the $e^{\pm}$ pairs created by the pair cascades.  These particles can be accelerated by the unscreened $\bm{E}_{\|}$ to high energies and produce the multi-wavelength radiation photons by the synchrotron, curvature, and inverse-Compton scattering processes. These photons have a back-reaction onto the particle motion and produce a radiative friction against the motion. The radiative friction brakes the particles in a direction opposite to their motion. We can expect that the particle acceleration and radiation can reach a stationary balance in the magnetosphere, which is called Aristotelian electrodynamics or radiation reaction limit. The balance condition between the Lorentz force and radiative friction gives \citep{mes99,pet16a,pet23}
\begin{eqnarray}
q(\bm{E}+\bm{v}\times\bm{B})=K\bm{v}.
\label{Eq18}
\end{eqnarray}
Where $K$ is a positive constant parameter. By taking the cross product to equation (\ref{Eq18}) with $\bm{B}$ and using a vector calculus, we can obtain the particle velocity
\begin{eqnarray}
\left(B^2+\frac{K^2}{q^2}\right)\bm{v}=\frac{K}{q}\bm{E}+\bm{E}\times\bm{B}+\frac{q}{K}(\bm{E}\cdot\bm{B})\bm{B}.
\label{Eq19}
\end{eqnarray}
By taking the dot product to Equation (\ref{Eq19}) with $\bm{v}$ and using a vector calculus, Equation (\ref{Eq19}) becomes
\begin{eqnarray}
K^4v^2-q^2(E^2-v^2B^2)K^2-q^4(\bm{E}\cdot\bm{B})^2=0.
\label{Eq20}
\end{eqnarray}
Assuming that the  particle velocity is equal to the light velocity ($v=1$), we have
\begin{eqnarray}
K^2=\frac{q^2}{2}\left[ E^2-B^2\pm\sqrt{(E^2-B^2)^2+4(\bm{E}\cdot\bm{B})^2} \right].
\label{Eq21}
\end{eqnarray}
By using the Lorentz invariants (\ref{Eq13}), we  find that the constant  $K$ is the solution of the following equations
\begin{equation}
\begin{split}
E^2-B^2&=K^2/q^2-B^2_0\;,\\
\bm{E}\cdot\bm{B}&=K\,B_0/|q|\;.
\end{split}
\label{Eq22}
\end{equation}
Substituting Equation (\ref{Eq22}) into Equation (\ref{Eq19}), we can retrieve the AE velocity expression \citep{fin86,gru12,gru13}
\begin{eqnarray}
{\bm v_{\pm}}=  {{\bm E} \times {\bm B}\pm(B_0{\bm {B}}+E_0{\bm {E}}) \over B^2+E^2_{0}},
\label{Eq23}
\end{eqnarray}
where the two signs correspond to positrons and electrons, they follow a different trajectory that only depends on the local electromagnetic fields in the magnetosphere. The AE current density can be derived by Equation (\ref{Eq23})
\begin{eqnarray}
{\bm J }=   {\rho_e {\bm E} \times {\bm B} + \rho_0 (B_0{\bm {B}}+E_0{\bm {E}}) \over B^2+E^2_{0}},
\label{Eq24}
\end{eqnarray}
where $\rho_0=\rho_{+}+\rho_{-}$ is the total charge density. The AE current density can by defined by giving a description for the total charge density $\rho_0$. However, there is no unique prescription for the current density since we do not know  the exact expression of  the total charge density. A formulation of the AE current density can be defined by introducing a pair multiplicity $\kappa$ \citep{cao20,pet20b}
\begin{eqnarray}
{\bm J }=  \rho_e  \frac {{\bm E} \times {\bm B}}{B^2+E^2_{0}}+ (1+\kappa)\left|\rho_e\right| \frac{ (B_0{\bm {B}}+E_0{\bm {E}}) }{ B^2+E^2_{0}},
\label{Eq25}
\end{eqnarray}
where $|\rho_e|$ can be understood as the charge density of the primary electrons, $\kappa|\,\rho_e|$ can be understood as  the  charge density of the secondary pairs from the pair cascades. Therefore, the pair multiplicity $\kappa$ is  physically related with the pair cascade processes in the magnetosphere. When $\kappa=0$, the AE current density is exactly consistent with the one given by \citep{gru13}.

\begin{figure*}
\center
\begin{tabular}{cccccccccccccc}
\includegraphics[width=16. cm,height=10. cm]{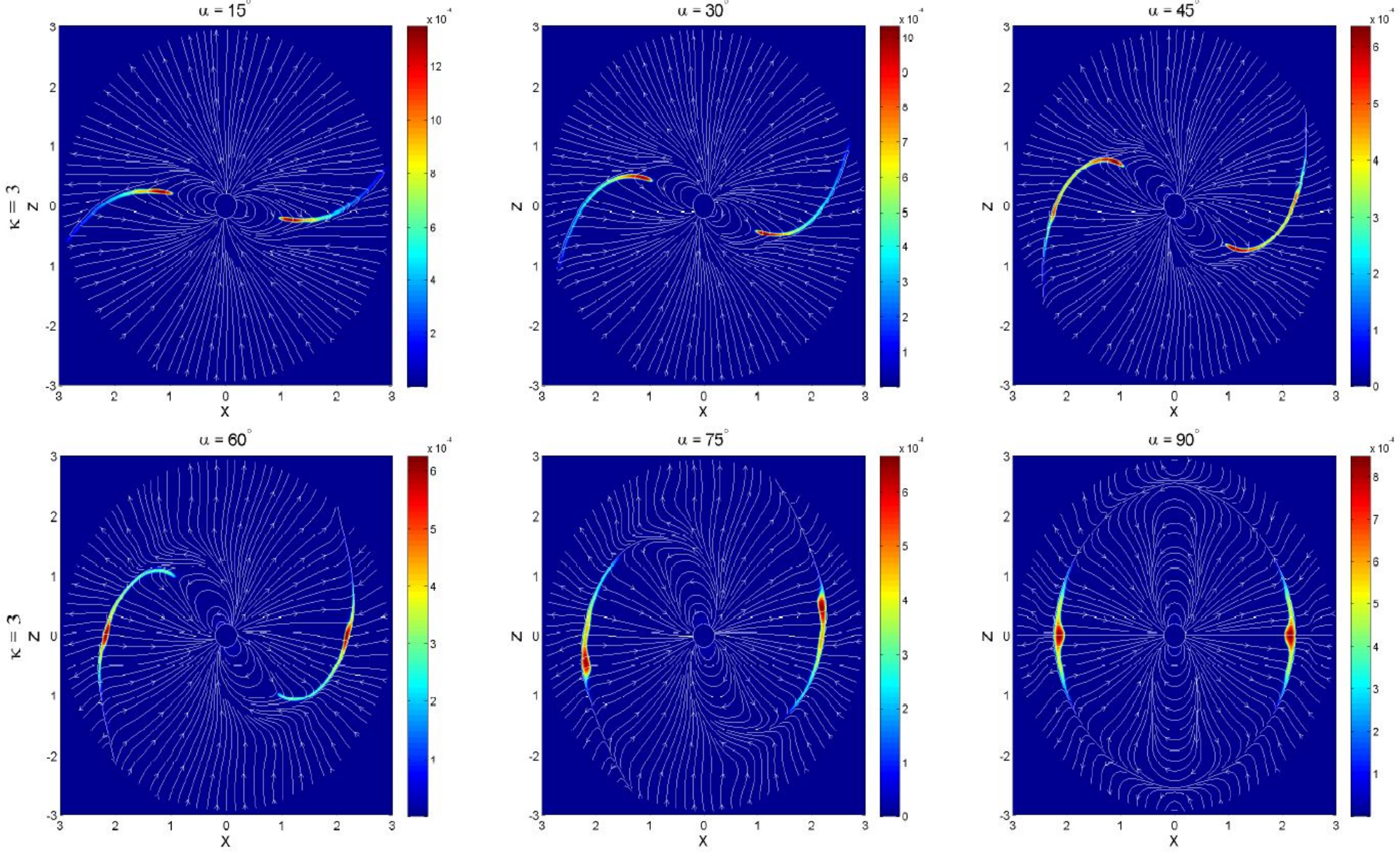}
\end{tabular}
\caption{The distributions of the magnetic field lines and accelerating electric fields in the combined force-free and AE magnetospheres for different inclined angles with the pair multiplicity $\kappa=3$ at the x$-$z plane. The figure is taken from Cao \& Yang (2020) \citep{cao20}.}
\label{fig14}
\end{figure*}

The resistive model can produce the local dissipations by relaxing the force-free condition, but they can not self-consistently include the backreaction of emission onto particle dynamics. The PIC simulation can model the pulsar magnetospheres by including the self-consistent feedback between the particle motions, the radiating photons, and the electromagnetic fields, but they cannot use the real pulsar parameters to model the pulsar magnetospheres due to the large ratio between the macroscopic scales to the microscopic scales. The combined force-free and AE model is a good approximation between the resistive and PIC models, which can include the back-reaction of the emitting photons onto particle dynamics and allows for the local dissipations  to produce the observed pulsar emission with the real pulsar parameters. A clever method with the combined force-free and AE was then developed to model the pulsar magnetosphere by using the force-free description where $E \leq B$ and the AE description where $E > B$. The combined force-free and AE magnetospheres for the oblique rotator in the limit of $\kappa=0$ was first computed by the finite-difference method \citep{con16a}. This study  was extended by introducing a new pair multiplicity and  the combined force-free and AE solutions  for the aligned rotator are simulated by the spectral method \citep{pet20a}. The combined force-free and AE solutions of the oblique rotator with a prescribed pair multiplicity  were first obtained  by Cao \& Yang (2020) \citep{cao20} and then  by P\'{e}tri (2022) \citep{pet22}, who both used the 3D spectra method to numerically solve the time-dependent Maxwell equations (\ref{Eq1}) with the AE current density (\ref{Eq25}). The high-resolution simulations of the combined force-free and AE magnetosphere for the oblique rotator were also explored to accurately resolve the $\bm{E}_{\|}$ regions in the current sheet  by the 3D spectral method \citep{cao22}. All these time-dependent simulations obtain a family  of the similar field  structures with the $\bm{E}_{\|}$ regions near the current sheet outside the LC for a range of the $\kappa$ values. As the pair multiplicity $\kappa$ increases, the combined force-free and AE magnetosphere tends to the force-free solution with the current sheet outside the LC,  and the $\bm{E}_{\|}$  region is more restricted to the current sheet outside the LC. The $\bm{E}_{\|}$ regions decrease with increasing $\kappa$ values and the $\bm{E}_{\|}$  region is only confined to the near the current sheet outside the LC for the high $\kappa$ values. Figure \ref{fig14} shows the distributions of the magnetic field lines and accelerating electric fields in the combined force-free and AE magnetospheres for different inclined angles with the pair multiplicity $\kappa=3$  at the x$-$z  plane. It is seen that all these solutions have a near force-free magnetosphere with the $\bm{E}_{\|}$ region only near the current sheet for all the inclination angles at the  high $\kappa$ value, which indicates that the particle acceleration and the high-energy emission mainly occur at the current sheet near the LC .

\begin{figure*}
\center
\begin{tabular}{cccccccccccccc}
\includegraphics[width=16. cm,height=10.5 cm]{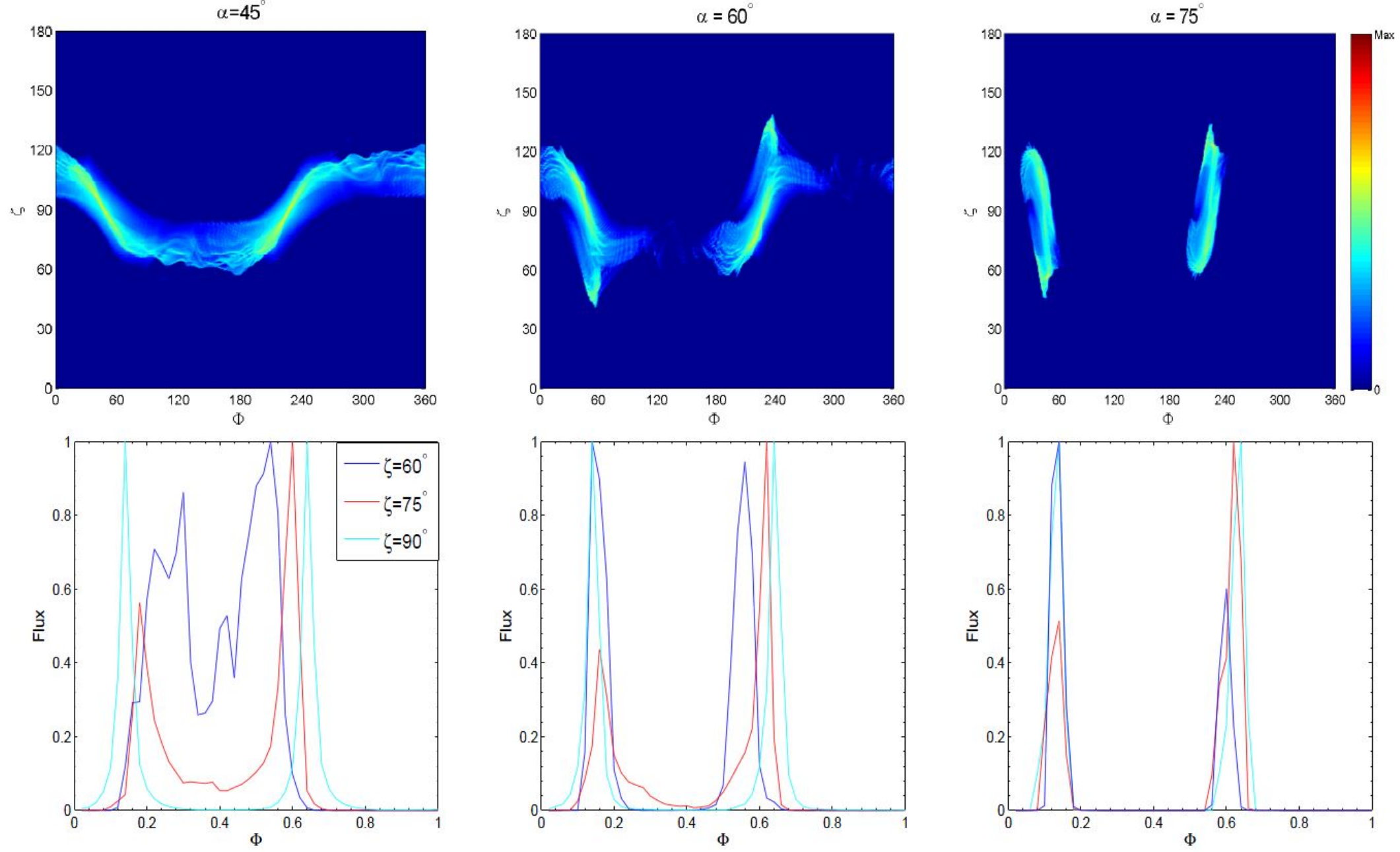}
\end{tabular}
\caption{The sky maps and the corresponding light curves at >1 GeV energies for a range of the inclination angles and view angles in the combined force-free and AE magnetospheres with the pair multiplicity $\kappa=3$. The figure is taken from Cao \& Yang (2022) \citep{cao22}. }
\label{fig15}
\end{figure*}

\subsection{Light curve and spectra modelling from the combined force-free and AE magnetospheres}

The combined force-free and AE magnetosphere provide a near force-free field structure with the $\bm{E}_{\|}$ distribution only near the current sheet outside the LC. We can use these  field structures and $\bm{E}_{\|}$ distributions to predict the pulsar $\gamma$-ray emission by integrating the particle trajectory with the AE velocity. The AE velocity expressions are extended to a new AE velocity expressions  by including the Landau-Lifshitz terms. It is found that the  new AE velocity expressions give the similar results in accuracy to the standard AE velocity expressions \citep{pet23,cai23a,cai23b}. It is also demonstrated that the AE velocity can correctly follow the particle trajectory in the strong electromagnetic field, which can be implemented to compute realistic pulsar light curves and spectra in the arbitrary pulsar magnetosphere \citep{pet19,pet20b}. The pulsar $\gamma$-ray light curves and spectra can be computed by the synchro-curvature radiation from each radiating particle \citep{zha97,vig15,tor18}
\begin{eqnarray}
\frac{d N_{sc}}{d E}=\frac{\sqrt{3}e^3\gamma y}{4\pi\hbar r_{\rm eff}}\left[(1+z)F(y)+(1-z)k_{2/3}(y)\right],
\label{Eq25}
\end{eqnarray}
with
\begin{eqnarray}
y=\frac{E}{E_c},\;\; E_c=\frac{3}{2}\hbar c Q_2 \gamma^3,\;\; z=(Q_2 r_{\rm eff})^{-2},
\label{Eq25}
\end{eqnarray}
\begin{eqnarray}
F(y)=y\int_{y}^{\infty}{K_{\rm 5/3}}(y)dy,
\label{Eq25}
\end{eqnarray}
\begin{eqnarray}
Q_2=\frac{\cos^2\psi}{\rho}\left(1+3\xi+\xi^2+\frac{r_g}{\rho}\right)^{1/2},
\label{Eq25}
\end{eqnarray}
\begin{eqnarray}
r_{\rm eff}=\frac{\rho}{\cos\psi}\left(1+\xi+\frac{r_g}{\rho}\right)^{-1},
\label{Eq25}
\end{eqnarray}
\begin{eqnarray}
r_{g}=\frac{m c^2 \gamma \sin\psi}{e B},\;\; \xi=\frac{\rho\sin^2\psi}{r_g\cos^2\psi },
\label{Eq25}
\end{eqnarray}
where $E$ is the photon energy, $E_c$ is the characteristic energy of the synchro-curvature photon, $\gamma$ is the particle Lorentz factor, $\rho$, $\psi$ and $r_g$ are the particle curvature radius, the pitch angle and the Larmor radius, respectively. The parameter $\xi$ determines the dominant synchro-curvature regime with  $\xi>>1$ at the synchrotron regime and
$\xi<<1$ at the curvature regime.

\begin{figure*}
\center
\begin{tabular}{cccc}
\includegraphics[width=16. cm,height=4.85 cm]{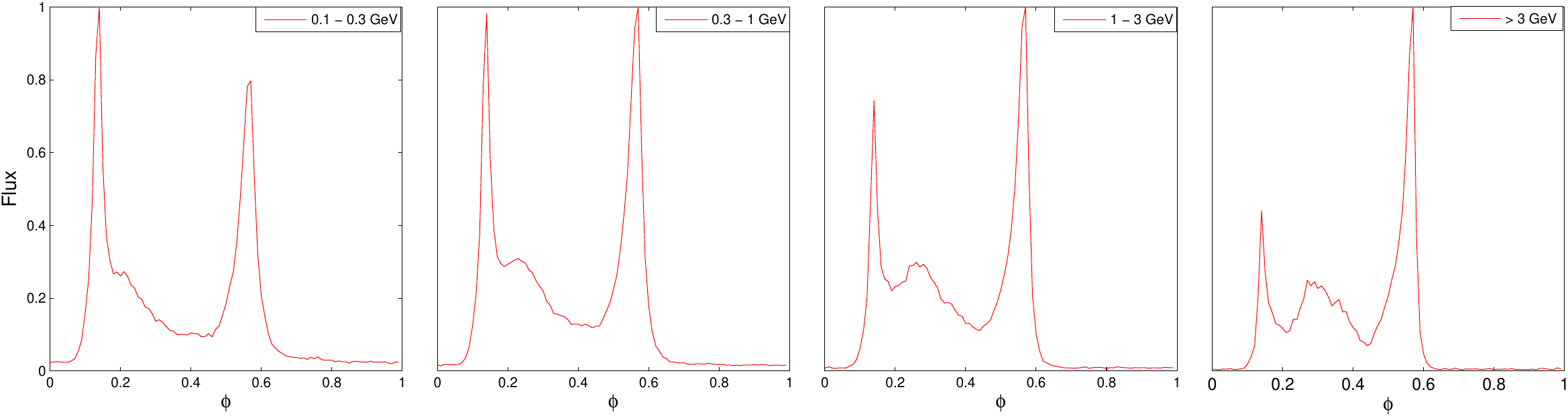}\\
\includegraphics[width=16. cm,height=4.85 cm]{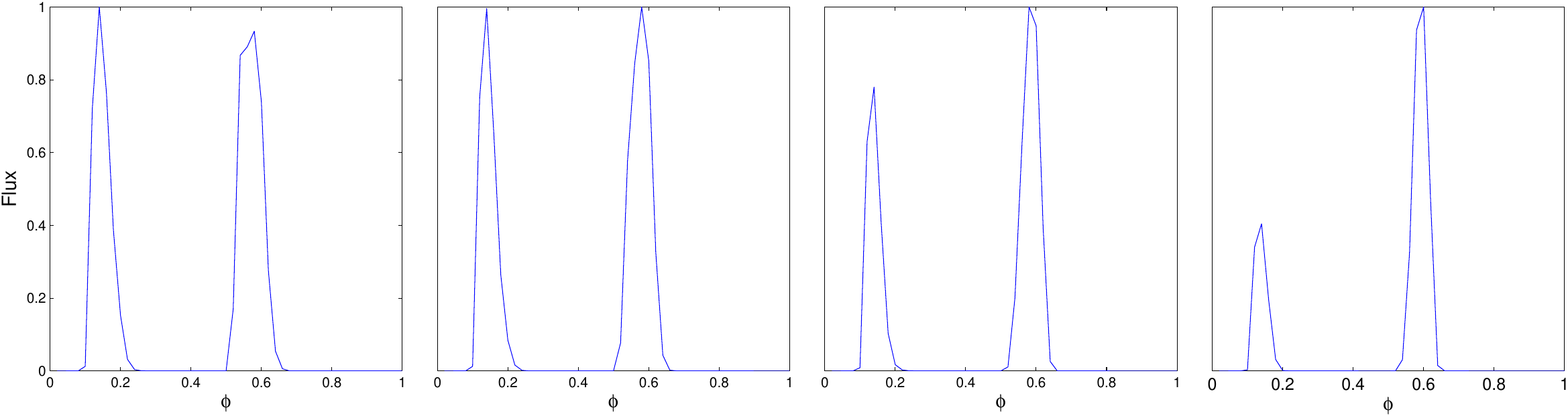}
\end{tabular}
\caption{A comparison of the predicted energy-dependent $\gamma$-ray light curves and the observed ones for the Vela pulsar.  The top panel is the observed $\gamma$-ray light curves taken from Abdo et al. (2013) \citep{abd13}. The bottom panel is the predicted energy-dependent $\gamma$-ray light curves in the combined force-free and AE magnetospheres. The figure is taken from Cao \& Yang (2024) \citep{cao24}. }
\label{fig16}
\end{figure*}

The properties of the pulsar $\gamma$-ray light curves and  spectra from the curvature radiation in the combined force-free and AE magnetosphere were first explored by Cao \& Yang (2022) \citep{cao22}, where the AE velocity was used to determine the particle trajectory and the Lorentz factor along each trajectory was computed by including the self-consistent $\bm{E}_{\|}$ values and the curvature radiation losses. When the direction of the photon emission along the direction of particle motion is assumed, the pulsar energy-dependent $\gamma$-ray light curves and spectra are produced by collecting the curvature photons from all radiating particle in sky maps.
Figure \ref{fig15} shows the sky maps and the corresponding light curves for different inclination angles and view angles in the combined force-free and AE magnetospheres with the pair multiplicity $\kappa=3$. The sky maps from the combined force-free and AE magnetosphere show the characteristics of the caustic emission from the current sheet, which are similar to those of the PIC model and  the other current sheet emission model. The double-peak $\gamma$-ray light curves can generally be produced for a broad range of inclination angles and view angles, and the  phase lags from the magnetic pole are in good agreement with the Fermi observation for the high pair multiplicity. The predicted $\gamma$-ray spectra can reach the  $\gamma$-ray energies with the cutoff energies around several GeV in agreement with those of the Fermi $\gamma$-ray spectra, which do not need the scale-up method as in the PIC simulations since they can be computed for realistic pulsar parameters.
The pulsar $\gamma$-ray  light curves and  spectra are further explored by performing a direct comparison with the Fermi observation for the Vela pulsar \citep{cao24}. It is found that the predict energy-dependent $\gamma$-ray light curves can well reproduce the observed trend of the $\gamma$-ray light curves with the energy evolution for the Vela pulsar (Figure \ref{fig16}), the decreasing ratio of the first peak to the second one with increasing energy is explained as the systematically larger  curvature radius and the higher cutoff energy  in the second peak.  The predicted $\gamma$-ray spectra can also well explain the phase-averaged spectra and phase-resolved spectra of the first and second peaks for the Vela pulsar. The observed higher spectral flux and cutoff energy in the second peak are also well reproduced, because  the second peak  of the $\gamma$-ray light curves  survives at the higher energy than the first peak one.

\section{Summary}
\label{sec9}

The last decades have witnessed the significant advances in the pulsar $\gamma$-ray observations made by  Fermi Gamma-Ray Space Telescope and ground-based Cherenkov telescopes. These $\gamma$-ray observations have obtained the valuable observational data about the pulsar $\gamma$-ray light curves and spectra in the GeV$-$TeV range. The modeling of the Fermi $\gamma$-ray light curves and spectra locates the $\gamma$-ray emission region at the outer magnetosphere near the LC or beyond the LC. The TeV $\gamma$-ray data from ground-based Cherenkov telescopes provide another important constraints on the particle acceleration and emission mechanics in the pulsar magnetospheres.
In parallel to the advances in the pulsar $\gamma$-ray observations, the self-consistent modeling of the pulsar magnetosphere have also made the significant breakthroughs from the early vacuum model to the recent plasma-filled magnetospheres by the numerical simulations. The current numerical simulations confirmed the general picture that the typical pulsar magnetosphere has a near force-free field structure with the particle acceleration in the separatrix near the LC and the current sheet outside the LC, which can provide a good match to the recent high-energy $\gamma$-ray observation. However, there is disagreement over whether the $\gamma$-rays emission are produced by the synchrotron radiation or the curvature radiation from the current sheet near the LC. The light curve and spectral information  is not sufficient to distinguish between different emission models and emission mechanisms in the magnetospheres. This degeneracy can be broken by using the additional constraints from the future polarization measurements.
The modeling of the combined multi-wavelength light curves, spectra, and polarization would put a stronger constraint on the the location and geometry of the particle acceleration and emission region as well as the emission mechanism. Realistic PIC simulations are prohibited by the large ratio between the plasma frequency and the pulsar rotation frequency. A more clever idea is required to perform the realistic meaningful simulations of the pulsar magnetosphere and predict the realistic pulsar emission.

\vspace{6pt}
\authorcontributions{All authors have read and agreed to the published version of the manuscript.}

\funding{L.Z. is partially supported from the National Natural Science Foundation of China 12233006. G. C. is supported from the National Natural Science Foundation of China 12003026 and 12373045, and the Basic research Program of Yunnan Province 202001AU070070 and 202301AU070082.}

\conflictsofinterest{The author declares no conflict of interest.}


\reftitle{References}

\end{document}